\documentclass[aps,prd,amsmath,amssymb,amsfonts,onecolumn,nofootinbib,preprint]{revtex4}
\usepackage{graphicx}% Include figure files
\usepackage{dcolumn}% Align table columns on decimal point
\usepackage{bm}% bold math
\usepackage{amsmath}
\usepackage{amssymb}
\usepackage{color}

\newcommand{\be}{\begin{equation}}
\newcommand{\ee}{\end{equation}}
\newcommand{\bea}{\begin{eqnarray}}
\newcommand{\eea}{\end{eqnarray}}
\newcommand{\bem}{\left(\begin{matrix}}
\newcommand{\eem}{\end{matrix}\right)}

\begin{document}

\title{Confinement/deconfinement transition from symmetry breaking in gauge/gravity duality}

\author{Mihailo \v{C}ubrovi\'c}
\affiliation{Institute for Theoretical Physics, University of Cologne, Z\"{u}lpicher Strasse 77, D-50937, Cologne, Germany}

\begin{abstract}
We study the confinement/deconfinement transition in a strongly coupled system triggered by an independent symmetry-breaking quantum phase transition in gauge/gravity duality. The gravity dual is an Einstein-scalar-dilaton system with AdS near-boundary behavior and soft wall interior at zero scalar condensate. We study the cases of neutral and charged condensate separately. In the former case the condensation breaks the discrete $\mathbb{Z}_2$ symmetry while a charged condensate breaks the continuous $U(1)$ symmetry. After the condensation of the order parameter, the non-zero vacuum expectation value of the scalar couples to the dilaton, changing the soft wall geometry into a non-confining and anisotropically scale-invariant infrared metric. In other words, the formation of long-range order is immediately followed by the deconfinement transition and the two critical points coincide. The confined phase has a scale -- the confinement scale (energy gap) which vanishes in the deconfined case. Therefore, the breaking of the symmetry of the scalar ($\mathbb{Z}_2$ or $U(1)$) in turn restores the scaling symmetry in the system and neither phase has a higher overall symmetry than the other. When the scalar is charged the phase transition is continuous which goes against the Ginzburg-Landau theory where such transitions generically only occur discontinuously. This phenomenon has some commonalities with the scenario of deconfined criticality. The mechanism we have found has applications mainly in effective field theories such as quantum magnetic systems. We briefly discuss these applications and the relation to real-world systems.
\end{abstract}

\maketitle

\section{Introduction}

% (1) Remarks from the printed copy+++
% (2) Explain better the groups G and H in Intro, define what is meant by gauge group and why U(1) is global etc+++
% (3) Fig. 6 compute a few more points+++
% (4) Why both T=0 and T>0 phase transition is BKT? Holography examples?
% (5) Everything after page 30+++

The gauge/gravity duality, AdS/CFT correspondence or holography \cite{w,gpk} is by now a well-established area, providing insights into fundamental issues of string theory and quantum gravity but also into strongly-coupled physics in various areas such as quantum chromodynamics (QCD) and condensed matter systems \cite{janbook}. In such studies, the spacetime has anti de Sitter (AdS) geometry at large distances, near the boundary of the space, while the interior is deformed away from AdS by various matter and gauge fields. This means that the high-energy behavior (ultraviolet, UV) of the field theory, determined by the near-boundary geometry, is conformally invariant but the interesting low-energy (infrared, IR) physics is determined by the geometry of the interior which can look differently for various configurations of fields and matter. The basic idea is that the radial coordinate on the gravity side corresponds to the energy scale in field theory: as we travel from the boundary toward the interior, we probe lower and lower energy scales.

One outstanding problem where AdS/CFT has provided some insights is the confinement/deconfinement transition in strongly coupled gauge theories. In the confined phase, only gauge-neutral bound states (mesons or baryons) can be observed. In the deconfined phase, individual gauge-charged particles are also observable. The fact that the gauge-charged excitations confine to form gauge-neutral bound states means that a gap opens, as we only see the gauge-neutral bound states at finite energies; the number of the degrees of freedom is effectively reduced at low energies. In AdS/CFT, this in turn means that the scale of the spacetime in the dual gravity model shrinks to zero in the interior. Such geometries are called soft-wall geometries, if the scale shrinks continuously, or hard-wall geometries if the spacetime is sharply cut off at some finite radius. Soft-wall geometries (which are more realistic than the hard-wall idealization) are obtained by coupling a neutral scalar -- dilaton -- to the metric in a non-minimal way. They were first used in so-called AdS/QCD studies in \cite{wconf,soft,umut}.

Typically, as the temperature rises, the system undergoes a confinement/deconfinement phase transition: when the system deconfines, the free energy of individual gauge-charged particles becomes finite, and they can be observed. This is the dominant mechanism in quark-gluon plasmas in QCD. But confinement/deconfinement is also present in condensed matter systems. Here the gauge field is not microscopic but emergent in the low-energy description. In the confined phase the degrees of freedom are bound into gauge-neutral excitations which are seen as normal electrons, i.e. quasiparticles. In the deconfined regime, the excitations are gauge-charged and not observable by ordinary probes in experiment. This might explain some non-Fermi liquid materials \cite{heavy,deconfcrit}. This topic was addressed e.g in \cite{hartnollhuijse} as well as in a series of very general and systematic studies by Kiritsis and coworkers \cite{bigkir,charmkir,goutkir}. In such systems, it is realistic to assume that deconfinement can also happen as a quantum phase transition, at zero temperature, when some parameter is varied. Deconfined gauge theories in AdS/CFT often have full conformal symmetry (dual to AdS geometry) or at least some form of anisotropic scale covariance (with different scaling exponents along different coordinates) which arguably can be expected to hold at high energies for many realistic gauge theories \cite{wconf} while a confined theory has an explicit scale because the energy of gauge-neutral bound states or, equivalently, the position of the wall $z_w$ along the radial direction sets a scale. The confinement/deconfinement transition can thus be treated also as a symmetry-breaking transition, where scale covariance is lost. The scaling properties of dilaton spacetimes have recently become known as hyperscaling geometries \cite{newkir,hyperscal} and have attracted attention also independently of the confinement/deconfinement problems.

Another class of problems where strongly-coupled models are provided by AdS/CFT are the order/disorder quantum phase transitions where some field $O$ acquires a vacuum expectation value (VEV). A textbook example is the famous holographic superconductor \cite{hhh,hhhprl,holscrev} where a charged field condenses, breaking the $U(1)$ symmetry, similarly to the superconducting transition in metals. While many such systems are described by the Landau-Ginzburg paradigm, this paradigm fails in some strongly-coupled systems. Many variations of such models have been proposed \cite{salvio,yanyawendil,albert} where a dilaton is also present, or it is precisely the dilaton that condenses. The work \cite{salvio} in particular addresses the setup similar as in our study: a scalar which condenses in the presence of a separate dilaton (however, explicit calculation with backreaction on dilaton and geometry was not done to check if the confinement/deconfinement transition exists). This opens an alley to study the interplay of the two transitions, the confinement/deconfinement transition and the order/disorder transition.

Our idea is to explore the interplay of the two above phenomena: the confinement/deconfinement transition and the condensation of an order parameter. They might in principle be independent, or one might foster or hinder the other. Well-studied examples of holographic superconductors \cite{hhh} or superconductor-dilaton systems \cite{yanyawendil} suggest that order parameter condensation often makes the length scale in the interior decrease faster: a charged black AdS-Reissner-Nordstr\"om black hole, which in deep interior has the AdS${}_2$ geometry of finite radius, upon condensation turns into a Lifshitz spacetime whose length scale vanishes in the interior \cite{janbook,holscrev}; in \cite{hartnollhuijse} this was interpreted as turning a fractionalized non-Fermi liquid into a system closer to a Fermi liquid. On general grounds one also expects that an ordered system can be expressed in terms of fewer degrees of freedom (in terms of the fluctuations of the order parameter rather than all microscopic degrees of freedom).

We will however present an example where the opposite occurs, i.e. the formation of a condensate destroys the soft-wall geometry and deconfinement takes place. Thus the phase transition is not a straightforward symmetry-breaking transition: on one hand, the condensate breaks a symmetry, on the other hand, another symmetry is restored as the deconfinement happens. This happens because the confinement scale (the energy gap) vanishes so scale invariance is restored. Our main interest is how this transition looks and what is its nature. We find that the phase transition can be continuous, contrary to the prediction of the Ginzburg-Landau theory where such transitions (where the two phases have different symmetries neither of which is a subgroup of the other) can only occur through phase coexistence or a first-order transition.

This has some common logic with the deconfined criticality concept of \cite{deconfcrit}. Denote the full symmetry group of the non-soft-wall geometry by $\mathbb{G}_1$ and its subgroup which remains after confinement by $\mathbb{G}_2$. We do not know what exactly $\mathbb{G}_{1,2}$ are but generically $\mathbb{G}_1$ will contain some scale invariance which stems from the scaling behavior of the IR geometry, while the confined system has a scale (the confinement gap) and thus $\mathbb{G}_2$ does not contain any scaling symmetry. Denote further the symmetry group broken by the condensate formation by $\mathbb{H}_1$ and its residual subgroup by $\mathbb{H}_2$ (in our case, we have $(\mathbb{H}_1,\mathbb{H}_2)=(\mathbb{Z}_2,\mathbb{I})$ for the neutral scalar and $(\mathbb{H}_1,\mathbb{H}_2)=(U(1),\mathbb{I})$ for the charged scalar. Now in our paper we have a transition from the confined-disordered phase (symmetry group $\mathbb{G}_2\otimes\mathbb{H}_1$) to the deconfined-ordered phase with the symmetry $\mathbb{G}_1\otimes\mathbb{H}_2$. We have $\mathbb{G}_2<\mathbb{G}_1$ and $\mathbb{H}_2<\mathbb{H}_1$ so the critical point partly breaks, and partly restores symmetry. The same situation occurs in deconfined criticality scenario but the detailed physics is different: at a deconfined critical point (\emph{only} at the critical point) there is an additional topological conserved quantity which governs the transition. We will comment on this in the paper in more detail, however no direct relation or equivalence can be established at this level. We cite the deconfined criticality as an inspiration and possible direction of future work, not something that our present results are directly relevant for.

Although the specific problem of how condensation of a scalar may influence the confinement/deconfinement transition was not studied so far to the best of our knowledge, a lot of work was done on Einsten-{Maxwell}-dilaton systems in other contexts. After the pioneering work in \cite{wconf} which first drew attention to the AdS/QCD alley of research, confinement was studied in \cite{soft} and more systematically in \cite{umutkir1,umutkir2} with finite temperature behavior further explored in \cite{umut}. These authors have studied a neutral Einstein-dilaton system and have classified geometries which lead to confinement as well as the nature of the phase transition (first-order or continuous). Charged systems have been studied in \cite{bigkir,goutkir,charmkir,newkir}. The non-condensed phases of our system (without the order parameter) are just a small subset of the systems studied in \cite{bigkir} and we will frequently compare our case to their general results throughout the paper. Charged EMD systems are particularly well studied as top-down constructions regularly include charged fields. The charged case we consider is also closely related to the dilatonic charged black holes considered in \cite{gubser1,gubser2} as possible candidates for gravity duals of Fermi liquids. The issue of a scalar condensation in the presence of dilaton is also rather extensively studied, e.g. in \cite{albert} but in these cases the dilaton does not lead to a soft wall geometry so there is no confinement which can be destroyed upon condensation. In \cite{port1,yanyawendil} the dilaton itself is charged (i.e., a charged scalar is coupled to the curvature) and the phenomenology was found to be similar to the basic holographic superconductor \cite{hhh}.

In section 2 we give the gravity setup and explain our model. Section 3 sums up different solutions for the geometry, depending on the bulk mass (conformal dimensions) of the scalar field and classifies the solutions into confined and deconfined ones. In section 4 we explain how the condensation of the order parameter proceeds and how it leads to deconfinement, and finally construct the phase diagram of the system. In the fifth section we study the response functions (conductivity, charge susceptibility and the retarded propagator of the scalar field) and show how various phases and their symmetries can be inspected from the response functions which are in principle measurable quantities. The last section sums up the conclusions and discusses possible directions of further work. The Appendix contains a detailed description of the numerical calculations.

\section{Gravity setup}

We have the Einstein-(Maxwell)-scalar-dilaton system in asymptotically AdS${}_{D+1}$ spacetime, with or without the Maxwell sector: the metric $g_{\mu\nu}$, the dilaton scalar $\Phi$ and the (neutral or charged) scalar $\chi$. If the system is charged, there is also the electric field $A_\mu$ where only the electric component $A_0$ is nonzero (we do not consider magnetic systems). The dilaton $\Phi$ couples to the curvature $R$ in the string frame and is always neutral; thus unlike the models where the dilaton (actually, a non-minimally coupled scalar) is itself the charged field that condenses, we want the dilaton to perform its usual work, i.e. to control the scale (and confinement). This will be crucial to study the influence of the order parameter on the confining properties. We find it more convenient to work in the Einstein frame where the dilaton does not couple non-minimally to the metric but to the matter fields only. The scalar $\chi$ is minimally coupled to gravity and to the Maxwell field with charge $q$ (including the possibility $q=0$). We now have the action:
\bea
\label{action0}&S&=\int dt\int d^Dx\sqrt{-g}(R-\Lambda+L_\Phi+L_\psi+L_{EM})\\
\label{action1}&L_\Phi&=-\xi\left(\partial\Phi\right)^2-V(\Phi)\\
\label{action2}&L_\chi&=-\frac{1}{2}Z(\Phi)(D\chi)^2-\frac{m_\chi^2}{2}\chi^2=-\frac{1}{2}Z(\Phi)(\partial\chi)^2-\frac{q^2}{2f(z)^2}Z(\Phi)A_0^2\chi^2-\frac{e^{-2A}}{2\xi f}m_\chi^2\chi^2\\
\label{action3}&L_{EM}&=-\frac{1}{4}\mathcal{T}(\Phi)F^2=-\frac{1}{2}\mathcal{T}(\Phi)(\partial A_0)^2.
\eea
This is just the minimal symmetry-allowed action for these fields apart from the exponential couplings of the dilaton. In string theory we would have $\xi=4/(D-1)$ but since our model is purely phenomenological we can leave it as an arbitrary positive constant. We have subtracted the constant piece, i.e. the cosmological constant $\Lambda=-D(D-1)/2$ from the dilaton potential, so the AdS solution corresponds to $\Phi=0$ (and $\chi=0$). The geometry is AdS${}_{D+1}$ in the far field (UV, near-boundary) region while, with a suitable choice of $V(\Phi),Z(\Phi),\mathcal{T}(\Phi)$ it narrows into a soft wall in the interior (IR). The AdS radius is rescaled to $L=1$. The potential of the scalar is fixed to just the mass term, like in \cite{hhh,hhhprl}, as it suffices to achieve condensation (and is a consistent truncation of more elaborate, top-down potentials). As explained in \cite{hhh}, the field $\chi$, even when charged, can be made real, i.e. its phase can be put to zero. %The charge-neutral case (in absence of dilaton) is also familiar \cite{hhhprl,holscrev} and was briefly considered as a special case also in the original paper \cite{hhh}.

Now we come to the question of choosing the model, i.e. the dilaton potentials $V(\Phi),Z(\Phi),\mathcal{T}(\Phi)$. The basic picture of confinement in AdS/CFT means the dilaton potential should produce a soft-wall geometry but we also want to study its interplay with the establishment of (bosonic) long-range order and condensation. We want to engineer the dilaton potentials so that the scalar is unstable to condensation into a hairy black hole with $\chi(z)\neq 0$ for some $m_\chi^2$ in the soft wall background \emph{and} that the soft wall dilaton is in turn unstable to transition into a non-confining (non-soft-wall) solution upon the formation of scalar hair. This means that, upon dialing $m_\chi^2$, we should have two possible solutions for the scalar, $\chi(z)=0$ and $\chi(z)\neq 0$, each with a different nonzero solution for the dilaton $\Phi(z)$. The following potentials will serve us well:
\bea
\label{dilpotv}V(\Phi)&=&V_0\Phi^{\frac{2\nu-2}{\nu}}e^{2\Phi}\\
\label{dilpotz}Z(\Phi)&=&Z_0e^{\gamma\Phi},~~D-1<\gamma<2D\\
\label{dilpott}\mathcal{T}(\Phi)&=&T_0e^{\tau\Phi},~~\tau>2D-4,~~\tau^2>\left(\gamma+\frac{D-2}{8}\right)^2+\frac{1}{D}.
\eea
The limitations for $\gamma$ and $\tau$ follow from the requirement that the stress-energy tensor of the EM field and also of the charged scalar field $\chi$ should stay finite and not dominate over the components of the Einstein tensor. In top-down constructions from supergravity the functions $Z(\Phi),V(\Phi),\mathcal{T}(\Phi)$ are typically all purely exponential in $\Phi$ (or linear combinations of such exponentials), with fixed exponent values. In our bottom-up approach these exponents are free parameters and by tuning these we can study the behavior we are looking for. We have added a power-law prefactor to (\ref{dilpotv}) for reasons of better analytical tractability: the soft wall solution for the scale factor $A(z)$ is simplified with this choice for $V(\Phi)$ and at the leading order reads just $A(z)=z^\nu$ with subleading corrections for $z\to\infty$ whereas with a purely exponential $V(\Phi)$ it would have be more complicated also at leading order. We conjecture that the phase diagram and the overall behavior of the system would be similar for $V\propto e^{\kappa\Phi}$. In a companion publication we derive our model from a superpotential which demonstrates the stability of the system, giving legitimacy to (\ref{dilpotv}). The prefactors $Z_0,\mathcal{T}_0$ merely rescale the amplitudes of $\chi,A_0$ and can be put to unity (they have no physical meaning). Notice the case $\nu=1$ is special: then we get the linear dilaton theory, the potentials $V,Z,\mathcal{T}$ become purely exponential and can be embedded in a supergravity action. Finally, the potentials (\ref{dilpotv}-\ref{dilpott}) are the expressions in IR: near the AdS boundary they are corrected to ensure the AdS asymptotics.

For analytical considerations it is convenient to parametrize the metric as:\footnote{In numerical calculations we find it convenient to use a different parametrization of the metric. Equations of motion and the description of the numerical algorithm can be found in Appendix A.}
\be
\label{metricir}ds^2=e^{-2A(z)}\left(-f(z)dt^2+\frac{dz^2}{f(z)}+d\mathbf{x}^2\right),
\ee
with the coordinates $(t,z,x_1,\ldots x_{D-1})$, where $x_i$ are the transverse spatial coordinates, i.e. the spatial coordinates in field theory and $z$ is the radial distance in AdS space: the AdS boundary (UV of the field theory) sits at $z=0$ and the interior (IR in field theory) is at $z\to\infty$. At equilibrium, the fields are static, homogenous and isotropic, so they depend only on $z$. The equations of motion read:
\bea
\label{einst1ir}&A''&+(A')^2=\frac{1}{D-1}\frac{1}{f^2}T_{00}+T_{zz}=\frac{1}{2(D-1)}Z(\chi')^2+\frac{1}{D-1}\xi(\Phi')^2\\
\label{einst2ir}&f''&-(D-1)f'A'=2\left(\frac{1}{f}T_{00}+T_{ii}\right)=2e^{2A}\mathcal{T}(A_0')^2\\
\label{dilatoneom}&\Phi''&+\left(\frac{f'}{f}-(D-1)A'\right)\Phi'-\frac{e^{-2A}\partial_\Phi V}{\xi f}-\frac{e^{2A}f}{2\xi}(\chi')^2\partial_\Phi Z-\frac{e^{-3A}}{f}(A'_0)^2\partial_\Phi\mathcal{T}=0\\
\label{scalareom}&\chi''&+\left(\frac{f'}{f}-(D-1)A'+\Phi'\frac{\partial_\Phi Z}{Z}\right)\chi'-\frac{2e^{-2A}}{2f}m_\chi^2\chi+\frac{q^2}{f^2}ZA_0^2\chi=0\\
\label{emeom}&A_0''&-\left((D-3)A'-\frac{\partial_\Phi\mathcal{T}}{\mathcal{T}}\right)A_0'-\frac{2Z}{f\mathcal{T}}e^{-3A}\chi^2A_0=0.
\eea
The prime denotes the radial derivative. As we have only two independent functions in the metric, it suffices to take two combinations of the Einstein equations. Due to homogeneity we have $T_{x_1x_1}=T_{x_2x_2}=\ldots=T_{x_{D-1}x_{D-1}}\equiv T_{ii}$ and the off-diagonal components are zero. The energy-momentum tensor $T^{\mu\nu}=T^{\mu\nu}_\Phi+T^{\mu\nu}_\chi+T^{\mu\nu}_{EM}$ reads
\bea
\label{ttirPhi}&T^{00}_\Phi&=\xi g^{zz}(\Phi')^2-V,~~T^{zz}_\Phi=T^{ii}_\Phi=-\xi g^{zz}(\Phi')^2-V\\
\label{ttirchi}&T^{00}_\chi&=\frac{Zg^{zz}(\chi')^2}{2}+\frac{Zg^{zz}A_0^2\chi^2}{2}-m_\chi^2\chi^2,~~T^{zz}_{\chi}=T^{ii}_{\chi}=-\frac{Zg^{zz}(\chi')^2}{2}-\frac{Zg^{zz}A_0^2\chi^2}{2}-m_\chi^2\chi^2\\
\label{ttirEM}&T^{00}_{EM}&=T^{zz}_{EM}=-\mathcal{T}g^{00}g^{zz}(A'_0)^2,~~T^{ii}_{EM}=-2\mathcal{T}g^{00}g^{zz}(A'_0)^2
\eea
In order to have AdS asymptotics, the metric functions must satisfy $A(z\to 0)=\log z$ and $f(z\to 0)=1$. The near-boundary expansion of the gauge field is of the form
\be
A_0(z\to 0)=\mu-\rho z^{D-2}+\ldots
\ee
which determines the chemical potential $\mu$ and the charge density $\rho$. One can work either in the canonical ensemble (fixing $\rho$) or in the grand canonical ensemble (fixing $\mu$). For our purposes it doesn't matter much which variant we choose; in the concrete numerical examples we always fix the chemical potential. The scalar has the near-boundary behavior:
\be
\label{chibnd}\chi=\chi_-z^{\Delta_-}(1+c_{-1}z+c_{-2}z^2+\ldots)+\chi_+z^{\Delta_+}(1+c_{+1}z+c_{+2}z^2+\ldots)
\ee
where the leading and subleading branches $\chi_\mp$ have the conformal dimension $\Delta_\pm=D/2\pm\sqrt{D^2/4+m_\chi^2}$. In field theory, one of these is the source of the order parameter $O_\chi$ dual to $\chi$ and the other is its the vacuum expectation value (VEV). We pick $\chi_+$ as the VEV, so the formation of the condensate means $\chi_+\neq 0$ for $\chi_-=0$ -- nonzero subleading component (VEV) for zero leading (source) term. It usually turns out that the scalar can condense for negative enough mass squared, i.e. for $m_\chi^2<m_{BF}^2$ for some bound $m_{BF}$ (Breitenlohner-Friedmann bound \cite{BFbound}) that depends on the spacetime, i.e. on geometry; in AdS${}_{D+1}$ of unit radius it is $m_{BF}^2=-D^2/4$. Similar asymptotics as in (\ref{chibnd}) hold for the dilaton $\Phi$ when the near-boundary form of the potential starts from a quadratic term: $V(\Phi(z\to 0))\sim m_\Phi^2\Phi^2+\ldots$. We tune $m_\Phi^2$ above the bound for condensation because we never consider the condensed state of the dilaton. This leaves $\Phi_-$ as the sole free parameter. Obviously, $\Phi_-$ sources some field theory operator $O_\Phi$ of dimension $D/2-\sqrt{D^2/4+m_\Phi^2}$ which does not condense and thus does not break a symmetry. Still, the value of $\Phi_-$ influences the bulk solution and consequently may influence the condensation of $\chi$ or the confinement/deconfinement transition. In accordance with the main idea of the paper, we mainly focus on the condensation of $O_\chi$ at fixed $\Phi$ and only briefly discuss the meaning of $O_\Phi$.

In absence of the scalar $\chi$ and apart from the subleading correction in the dilaton potential $V$, our system is one of the many cases of Einstein-dilaton and Einstein-Maxwell-dilaton systems considered systematically in \cite{bigkir}. Our parameter values are similar to a solution that the authors of \cite{bigkir} call "near-extremal case". For each solution, we check that the value of the parameters we use for $\gamma,\delta,\nu$ are consistent with the Gubser criterion for "good" curvature singularities in IR \cite{gubsing}. A good singularity means that, even though the curvature becomes infinite at $z\to\infty$, it can be trapped by a horizon. A systematic discussion of allowed parameter values (for purely exponential potentials) can be found in the cited reference \cite{bigkir}. The exponent $\nu$ is also a free parameter with the limitation $\nu\geq 1$. In numerical calculations, unless specified differently, we take $\nu=2$ and $D=4$ for calculations, though any $D>2$ again leads to similar results. An account of numerical calculations can be found in the Appendix; the procedure is essentially iterative, repeatedly computing the profile of the scalar $\chi(z)$ and then updating the metric and the dilaton in the presence of $\chi(z)$.

\section{Solutions in the infrared: soft-wall and AdS-like}

\subsection{Neutral solutions}

\subsubsection{No symmetry breaking}

At zero temperature (which is central for studying the ground state) the space extends to $z\to\infty$. The authors of \cite{umutkir2} have performed a classification of asymptotically AdS Einstein-dilaton systems (without other fields), motivated by AdS/QCD studies. Their results can be summed up as follows. The scale factor $A(z)$ either has a singularity at finite $z$, or at $z=\infty$. In the former case, the metric can be conformally equivalent to AdS with $A(z)\sim\alpha\log z$ (type Ib geometry), which is never confining whereas the soft-wall solutions with $A(z)\sim z^\nu$ (type Ia geometry) are confining for $\nu\geq 1$. If the singularity is to be found at finite $z=z_W$, then the logarithmic approach $A(z)\sim\log (z_W-z)$ (type IIb geometry) does not give confinement whereas a power-law $A(z)\sim 1/(z_W-z)^\nu$ (type IIa solution) does, for any $\nu$.\footnote{Let us quickly remind the reader where this comes from. The defining criterion for confinement is that the Wilson loop operator follows the area law. The Wilson loop, defined as the potential energy of a quark-antiquark pair separated by distance $L$, is holographically expressed as the action of a classical string embedded in spacetime, with a rectangular loop at the AdS boundary with sides equal to $L$ and the time $T$. If the metric is of the form (\ref{metricir}), one can plug it in into the expression for the string action and find the action scales as $e^{-2A(z_s)}$, where $z_s$ is a stationary point: $A'(z_s)=0$. From this the above conclusions follow, bearing in mind that one may have $z_s\to\infty$.} We have nothing to add here: our system is a special case of the systems considered in \cite{umutkir2}, with slightly different $V(\Phi)$.

To solve our equations of motion (\ref{einst1ir}-\ref{scalareom}), notice first that the equation (\ref{einst2ir}) is decoupled from all matter fields and yields the solution
\be
\label{fsol}f(z)=C_0+C_1\int dze^{(D-1)A(z)}.
\ee
A growing scale $A(z)$ in the interior would lead to a bad singularity according to the criterion of Gubser \cite{gubsing}.\footnote{To remind the reader, the paper \cite{gubsing} shows that a curvature singularity is physically meaningful if it can be obtained as the limit of a geometry with horizon, so that the horizon hides the singularity.} Therefore, we need to suppose that $A(z)$ is a monotonically growing function of $z$, as also discussed in \cite{umutkir2}. This in turn means that the non-constant term in (\ref{fsol}) is likewise growing, so $C_1<0$ (in order to have a solution for the position of the horizon, defined by $f(z_{hor})=0$) and for correct AdS asymptotics $C_0=1$. Now $C_1$ is determined by the boundary condition in the interior: at zero temperature, the space is infinite so $C_1=0, f(z)=1$ as expected for a neutral system. At nonzero temperature $T$, the position of the horizon is determined by the condition $f(z_h)=0$.

We are left with one Einstein equation for $A(z)$ and two Klein-Gordon-like equations for the two scalars. It is easiest to start from an ansatz $A(z)\sim z^\nu$ to get a soft wall (type Ia) solution
\bea
\nonumber &A(z)&=z^\nu\left(1+\frac{a_1}{z}+\frac{a_2}{z^2}+\ldots\right)\\
\nonumber &f(z)&=1-\frac{(D-1)^{1/\nu}}{\nu}\frac{\mathcal{M}}{z^{\nu-1}}e^{(D-1)z^\nu},~\chi(z)=0\\
\nonumber &\Phi(z)&=\sqrt{\frac{D-1}{\nu\xi}}\sqrt{\nu z^{2\nu}+(\nu-1)z^\nu}\left(1+\frac{\phi_{11}}{z}+\frac{\phi_{12}}{z^2}+\ldots\right)+\\
\label{geoIa}&+&\frac{\nu-1}{\nu\sqrt{\xi}}\log\left(\nu z^{\frac{\nu}{2}}+\sqrt{\nu^2z^\nu+\nu^2-\nu}\right)\left(1+\frac{\phi_{21}}{z}+\frac{\phi_{22}}{z^2}+\ldots\right).
\eea
These forms are exact as $z\to\infty$ and the coefficients $a_i,\phi_{ij}$ can be found analytically at arbitrary order in principle. We are not interested in the details of the small $z$ (UV) geometry, as long as enough free parameters remain that the solution can be continued to the AdS${}_{D+1}$ boundary conditions. The red shift function includes the rescaled black hole mass $\mathcal{M}$, which is related to the position of the horizon as $\mathcal{M}=z_h^{\nu-1}$, the horizon being determined through the transcendental equation $f(4\pi D/T)=0$ which we will not explore here in detail. Importantly, the thermal solution smoothly crosses into the zero temperature solution and all temperatures down to $T=0$ are defined, which is not always the case with Einstein-(Maxwell)-dilaton systems, see e.g. \cite{bigkir}. There is another solution, however: starting from the ansatz $A(z)\sim\alpha\log z$ we get a type Ib solution
\bea
\nonumber &A(z)&=\alpha\log z\left(1+\frac{a_1}{z}+\frac{a_2}{z^2}+\ldots\right),~~\alpha=\frac{\xi}{D-1+\xi}\\
\nonumber &f(z)&=1-\frac{\mathcal{M}}{\alpha(D-1)}z^{(D-1)\alpha},~\chi(z)=0\\
\label{geoIb}&\Phi(z)&=\phi_0\log z\left(1+\frac{\phi_1}{z}+\frac{\phi_2}{z\log z}+\frac{\phi_3}{z^2}+\frac{\phi_4}{z^2\log z}+\ldots\right),~~\phi_0=\frac{D-1}{D-1+\xi}.
\eea
Which of these is the ground state is to be determined by comparing the free energies, our task in the next section (it turns out the confining solution Ia is the correct choice). These solutions have a curvature singularity at $z\to\infty$: the Ricci scalar for (\ref{geoIa}) is
\be
\label{ricciia}R=-D(D-1)\nu^2e^{2z^\nu}z^{2\nu-2}+\ldots
\ee
which diverges for $z$ large but can be trapped by a thermal horizon for any finite $z_h$ so that $R$ is finite as $z_h\to 0$. This follows from the form of $f(z)$ in (\ref{geoIa}) and makes the solution physically meaningful.

\subsubsection{Symmetry-breaking order parameter}

Now consider the symmetry-broken solution with $\chi(z)\neq 0$. If we require the physically logical (and simplest) choice of purely exponential $Z(\Phi)$ as in (\ref{dilpotz}), then the only way to satisfy (\ref{einst1ir}) while keeping the scaling function $A(z)\sim z^\nu$ is to "reduce" the dilaton, i.e. make its growth slower than $z^\nu$: otherwise, an additional source on the rhs of equation (\ref{einst1ir}) can only make $A(z)$ grow even faster, never slower (remember the rhs is the kinetic energy of the scalar field which cannot be negative; adding a new nonzero field cannot reduce the sum). Thus we seek for a scalar $\chi(z)$ which, when coupled to $\Phi(z)$, gives it a logarithmic behavior $\Phi(z)\sim\phi_0\log z$. Such a solution indeed exists. We deliberately postpone the discussion of the mechanism of the scalar instability to condensation, i.e. of the scalar fluctuations in background (\ref{geoIa}) which lead to the new solution discussed in this subsection. This mechanism (and the value of $m_\chi^2$ at which it happens) will be discussed in the next section, before constructing the phase diagram. For now we are content to show that the solution exists. To the best of our knowledge, this kind of solution was not analytically constructed in earlier work.

The solution is now of type IbC (Ib with condensate):
\bea
\nonumber &A(z)&=\alpha\log z\left(1+\frac{a_1}{z}+\frac{a_2}{z^2}+\ldots\right),~~\alpha=\frac{\gamma+2}{2\gamma-2(D-1)}\\
%\nonumber f(z)=1-Ce^{(D-1)z^\nu}+\frac{2a_1^2}{2\nu-3}e^{(\tau+3-D)z^\nu}z^{3-2\nu}\\
\nonumber &f(z)&=1-\frac{\mathcal{M}}{\alpha(D-1)}z^{(D-1)\alpha}\\
\nonumber &\Phi(z)&=\phi_0\log z\left(1+\frac{b_1}{z}+\frac{b_2}{z\log z}+\frac{b_3}{z^2}+\frac{b_4}{z^2\log z}+\ldots\right),~~\phi_0=\frac{D+1}{\gamma+1-D}\\
\nonumber &\chi(z)&=\chi_0z^{-\frac{\gamma\phi_0}{2}},~~\phi_0=\frac{D+1}{2(\gamma+2)}\alpha=\frac{D+1}{\gamma-(D-1)}\\
\label{geoIbC}&\chi_0&=\frac{\sqrt{2(\gamma^2+2\gamma+2D^2(\gamma+2-2\xi)-4\xi-D((\gamma+2)^2+8\xi))}}{(D+1)\gamma}.
\eea
Interestingly, the value of $\nu$ does not appear in the solution at leading order (of course, it does appear in the subleading corrections $a_i,b_i$). The solutions for $\phi_0,\chi_0$ show that we need the condition $\gamma>D-1$ to avoid the growing metric scale in the interior. The crucial observation in the above discussion was that adding bosonic fields (for which $T_{00}/f^2+T_{zz}$ is always positive) cannot destroy the soft wall solution. We find there is no solution with two scalars, $\Phi$ and $\chi$, and with the couplings (\ref{dilpotv}-\ref{dilpotz}), which has a soft-wall metric scale behavior $A(z)\sim z^\nu$. This can be seen more rigorously from the superpotential approach. There is thus an interesting bifurcation-like behavior as the amplitude of the order parameter field is varied: there are two competing solutions for $\langle O_\chi\rangle=\chi(z=0)$ but only one of them survives as $\langle O_\chi\rangle$ grows away from zero. Is this solution acceptable? The curvature behaves as
\be
\label{ricciib}R=-4\alpha(3\alpha+2)z^{2\alpha-2}+\ldots\propto z^{\frac{\gamma-2D}{\gamma-D+1}},
\ee
the exponent being positive precisely in the allowed interval of $\gamma$ values, $D-1<\gamma<2D$. Thus we again have a singularity, and it is again a "good" singularity according to \cite{gubsing}. This is in line with the results of \cite{bigkir} for "near-extremal" solutions: acceptable solutions are only those with a singularity; those without a curvature singularity are cosmological solutions with an unacceptable singularity at small $z$.

\subsection{Charged solutions}

\subsubsection{No symmetry breaking}

Instead of a neutral scalar we now take a charged scalar, i.e. the typical holographic superconductor setting, coupled to a dilaton. The results should not depend crucially on the spin of the charged field as long as it is integer; half-integers fields, i.e. fermions may well behave differently as they have different pressure (spatial components of the stress tensor). We will not analyze the fermionic case here.

For further convenience we adopt the terminology of \cite{hartchap,hartnollhuijse}, used also in \cite{newkir}, to roughly classify the charged solutions in terms of the charge distribution in the bulk and how it influences the geometry. On one hand, we have (1) IR-neutral solutions where the \emph{Maxwell contribution} to $T_{\mu\nu}$ is subleading so that the IR geometry is not influenced by $A_0(z)$ in the first approximation, as opposed to (2) IR-charged solutions where $A_0(z)$ contributes at leading order. The second criterion is whether the solution is fractionalized or coherent: (a) fractionalized solutions are those where \emph{the charged fields} do not contribute to $T_{\mu\nu}$ and thus to geometry in the IR at leading order whereas in (b) cohesive solutions they contribute. In the fractionalized case the electric flux in the IR $\int\star\left[\mathcal{T}(\Phi)F\right]$ is non-zero while it is zero for cohesive solutions. The physical interpretation of the fractionalized/coherent dychotomy is still unclear. The logical explanation would be that in the fractionalized case the charge-carrying degrees of freedom are not those which are seen in the spectrum as they are charged under the gauge group and are not seen by the gauge-neutral probe ("gauginos"), as opposed to the gauge-neutral composite excitations of the coherent case ("mesinos"). This interpretation suggests a close relation between the confinement/deconfinement and coherence/fractionalization. The trouble is that many examples exist both of fractionalized but confined systems (the dilatonic black holes of \cite{gubser1,gubser2}) and coherent but deconfined systems (the electron star and the dilatonic electron star of \cite{hartnollhuijse}). While confinement is about the behavior of the Wilson operator and the gauge field excitations, coherence is about the emergence of stable composite gauge-neutral excitations. Examples where the quarks emerge only after the gauge field compactifies are known in AdS/CFT \cite{myers} but the understanding of the phenomenon is lacking. We plan to address this issue in more detail in future work; here we will just state the fractionalization/coherence nature of our geometries and comment briefly on the interpretation in the conclusions. For more information on the general problems of fractionalization in this context see \cite{newkir,gubser2}.

Let us now study the charged solutions. Notice first that a charged solution without condensate can only exist in the presence of a charged horizon. Such a solution must be fractionalized as none of the charge carriers have a dual VEV at the boundary. It reads
\bea
\nonumber &A(z)&=z^\nu\left(1+\frac{a_1}{z}+\frac{a_2}{z^2}+\ldots\right)\\
\nonumber &f(z)&=1-\mathcal{M}(T)e^{(D-1)z^\nu}+\frac{2Q^2}{2\nu-3}e^{(\tau+4-2D)z^\nu}z^{3-2\nu}\left(1+\frac{f_1}{z}+\frac{f_2}{z^2}+\ldots\right)\\
\nonumber &\Phi(z)&=z^\nu\left(1+\frac{\phi_1\log z}{z}+\frac{\phi_2}{z}+\ldots\right),~~\chi(z)=0\\
\label{geoIaQ} &A_0(z)&=a_0-Qe^{-(\tau-(D-3))z^\nu}z^{1-\nu}\left(1+\frac{a_1}{z}+\frac{a_2}{z^2}+\ldots\right).
\eea
Now the horizon carries the charge $Q$ and at zero temperature $\mathcal{M}(T=0)=0$, so the extremal horizon is degenerate and located at $z=\infty$. Again, we are not interested in the (complicated) analytical form of $\mathcal{M}(T)$. The electric flux at the horizon is $\sqrt{-g}g_{00}g_{zz}\mathcal{T}(\Phi)A_0'=\mathcal{T}(\Phi)A_0'\sim e^{-(D-3)z^\nu}\times e^{\tau z^\nu}a_1e^{(D-3-\tau)z^\nu}\sim a_1$ which is a generically nonzero constant for $z\to\infty$, meaning that the solution is fractionalized. On the other hand, it is confining, as it is of type Ia (we call it IaQ, as it has charge) and the metric scale diminishes exponentially in the IR (we call it IaQ to emphasize it is charged). In fact, this solution is quite similar to the top-down dilatonic black hole with two-exponent potential discussed in \cite{gubser1,gubser2}. Although fractionalized, it still confining so it fits into our main story: deconfinement from independent symmetry breaking.
%We get a confining, charged soft wall, a solution we deem IaQ (Ia charged):
%\bea
%\nonumber A(z)=z^\nu\left(1+\frac{a_1}{z}+\frac{a_2}{z^2}+\ldots\right)\\
%\nonumber f(z)=e^{-\left(\tau-\left(2D-4\right)\right)z^\nu}z^{1-\frac{1}{\nu^2}}\left(1+\frac{f_1}{z}+\frac{f_2}{z^2}+\ldots\right)\\
%\nonumber \Phi(z)=XXX\\
%\nonumber A_0(z)=a_0e^{-(3+\tau)z^\nu},~~a_0=-\frac{\nu^2}{(3+\tau)D}\rho_0\\
%\label{geoIaQ} \rho(z)=\rho_0\sqrt{f(z)}z^{2\nu-2},\rho_0=-\left(10+2D^2+\tau^2+7\tau-\left(3\tau+8\right)D-2\xi\right)\frac{\nu^2}{4},
%\eea
%Comparing the stress tensors by plugging in the solution, we easily find that $T_{EM}\ll T_\Phi$ for $z\to\infty$, so according to the criterion of \cite{bigkir} the solution is IR %neutral. XXX Now comment on the non-confining option. Does it still exist?

\subsubsection{Symmetry-breaking order parameter}

Postulating a nonzero profile for the scalar field and requiring that the scalar contributes at leading order in the equation (\ref{einst1ir}), we find the solution IbQC, the non-confining charged solution:
\bea\nonumber &A(z)&=\alpha\log z\left(1+\frac{a_1}{z}+\frac{a_2}{z^2}+\ldots\right)\\
\nonumber &f(z)&=1-\mathcal{M}(T)\frac{z^{(D-1)\alpha+1}}{(D-1)\alpha+1}+\frac{2Q^2}{z^\beta}\left(1+\frac{f_1}{z}+\frac{f_2}{z^2}+\ldots\right)\\
\nonumber &\Phi(z)&=\phi_0\log z\left(1+\frac{\phi_1\log z}{z}+\frac{\phi_2}{z}+\ldots\right)\\
\nonumber &\chi(z)&=\chi_0z^{-\frac{\gamma\phi_0}{2}}\left(1+\frac{\chi_1}{z}+\frac{\chi_2}{z^2}+\ldots\right)\\
\label{geoIbQC} &A_0(z)&=a_0-Qz^{-\frac{10+11\gamma+9\tau}{10+10\gamma+8\tau}}\left(1+\frac{a_1\log z}{z}+\frac{a_2}{z}+\ldots\right),
\eea
and the exponents read
\be
\alpha=\frac{4+4\gamma+3\tau}{5+5\gamma+4\tau},~~\beta=\frac{2\tau+3\gamma+2}{4\tau+5\gamma+5},~~\phi_0=\frac{1}{4\tau+5\gamma+5}.
\ee
The charged horizon is still degenerate at zero temperature. Comparing the stress tensors by plugging in the solution (\ref{geoIbQC}) into (\ref{ttirPhi}-\ref{ttirEM}), we easily find that $T_{EM}\ll T_\Phi,T_\chi$ for $z\to\infty$, so according to the criterion of \cite{bigkir} the solution is IR neutral. Being of type Ib (we denote it IbQC, as it is charged and has the condensate), it is not confining, and the IR flux is $z^{-2-\frac{9\gamma+5\tau+8}{10+10\gamma+8\tau}}$ which goes to zero for $z\to\infty$ since $\gamma$ and $\tau$ are positive and all the coefficients in both numerator and denominator of the exponent are positive. The solution IbQC is thus coherent and deconfined. On one hand, the fact that the non-condensed solution IaQ is fractionalized while the condensed solution IbQC is coherent is perfectly logical, since in the non-condensed case all the charge is on the horizon, whereas in the presence of the condensate it carries all the charge. On the other hand, the fact that the fractionalized solution is confined and the coherent one is deconfined may sound strange; e.g. in \cite{newkir} the intuition is expressed that confined solutions should be coherent. But as we have already commented the zoo of field theories in gauge/gravity duality offers many counterexamples. At least, one expects that the coherent nature of the systems shows up as poles, i.e. bound states in the bottom half-plane of complex-frequency response functions of matter fields (the scalar $\chi$), independently of the presence or absence of confinement. We will check this in Section V.

Can we get a soft wall with charged condensate? We were unable to find such a solution either analytically or numerically. The conclusion is again that the competition of two scalars (dilaton and order parameter) destroys the confining solution. Of course, by adjusting the potentials $V,Z,\mathcal{T}$ we could get many different phase diagrams but in the present model there is a strict competition between the soft wall and the condensate. Finally, the singularity properties of both charged solutions are analogous to the charge-neutral case: the singularities exist but are physically allowed.

\subsection{Resume of the geometries}

\begin{figure}[ht!]
\begin{center}
(A)\includegraphics[width=0.45\linewidth]{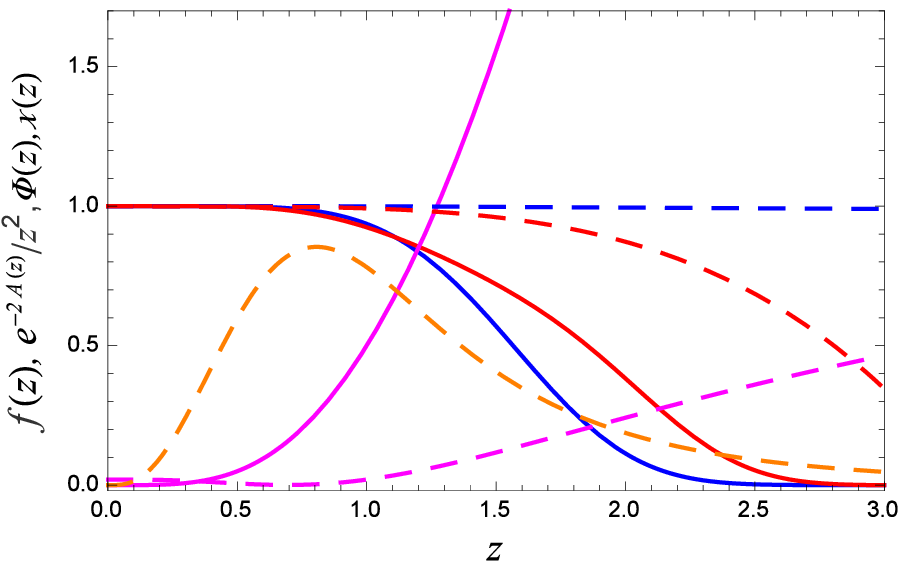}
(B)\includegraphics[width=0.45\linewidth]{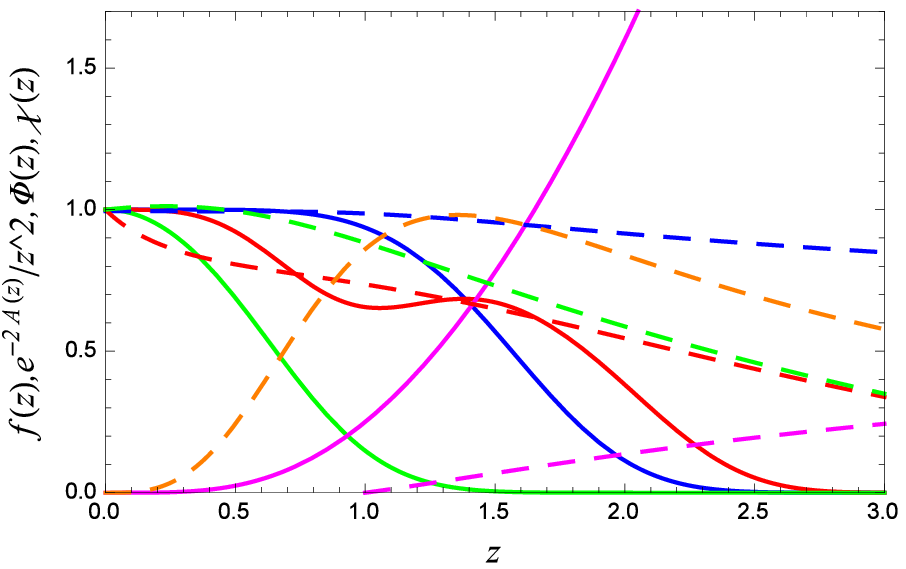}
\caption{(A) The metric functions $e^{-2A(z)}z^2,f(z)$ (blue, red) and the bulk fields $\Phi(z),\chi(z)$ (magenta, orange) in the confined regime (geometry Ia, full lines) and in the deconfined regime (geometry IbC, dashed lines). The blue line corresponds to the ratio of the scale factor in our system and the AdS scale factor $1/z^2$. The confining regime has a soft wall in the IR and its IR scale falls practically to zero already at $z\sim 3$. (B) Same as the previous figure but for the charged field at the chemical potential $\mu=1$; now we plot also the gauge field $A_0(z)$ (green). The basic phenomenology is the same as in (A): the soft wall broadens and the scale factor $e^{-2A}$ has no characteristic scale $z_W$ at which it falls off rapidly. The plots are in $D=4$ and the parameter values are $\nu=2$, $\gamma=4$ (both A and B), and $\tau=5$ for the charged case (B). For the neutral case we pick $m_\chi^2=1/4$ and $m_\chi^2=-1/4$ whereas for the charged case we have $m_\chi^2=8$ and $m_\chi^2=4$.}
\label{figgeom}
\end{center}
\end{figure}

We have found five solutions: Ia, Ib, IaQ, IbC, IbQC. Only two of them compete in the same regime, Ia and Ib, and the preferred solution has to be found by computing the energy. Geometries Ia, IaQ are confined whereas Ib, IbC, IbQC are deconfined. Among the charged geometries, IaQ is fractionalized whereas IbQC is coherent, and both are IR neutral. In Fig.~\ref{figgeom} we plot the metric functions $A(z),f(z)$ and the bulk profile of the dilaton and the scalar field $\Phi(z),\chi(z)$ at zero temperature, at zero chemical potential in the panel (A) and at finite chemical potential in the panel (B). The most obvious feature of the solutions is the sharp exponential fall-off of the scale factor $e^{-2A}$ for soft-wall geometries versus much slower fall-off for deconfined solutions where the blue curve $e^{-2A}z^2$ is almost flat, i.e. the solution behaves almost as AdS in the IR. This is logical, as the volume in the IR counts the degrees of freedom of the low-energy excitations; at low enough energies, such excitations are completely absent in the confined phase. In fact, as can be seen from the analytical form of the solutions (\ref{geoIbC},\ref{geoIbQC}), the factor $e^{-2A}$ in the deconfined phase behaves as a power law just like in AdS, only with a different power. In \cite{newkir,bigkir,hyperscal} such geometries are classified in terms of hyperscaling exponents, where the time, space and energy (i.e., radial distance in AdS) are each scale-covariant but with different exponents. It turns out these three exponents can be described by combinations of two parameters, the Lifshitz exponent $\zeta$ and the hyperscaling violation exponent $\theta$; if $\theta=0$ the geometry obeys the hyperscaling whereas for $\theta\neq 0$ it is hyperscaling-violating. For a Lorentz-invariant system we have $\zeta=1$; values different from unity mean that the dispersion relation is nonlinear and the Lorentz invariance broken. The neutral deconfined geometries (\ref{geoIb}) and (\ref{geoIbC}) have $\zeta=1$ but the hyperscaling exponent is nontrivial and reads $\theta=D(1+\alpha)$. The charged version (\ref{geoIbQC}) has both exponents nontrivial ($\zeta>0\neq 1$ and $\theta\neq 0$). Note that the $\zeta<0$ case is hard to interpret physically and thus we have checked that all of our geometries have $\zeta>0$.

\section{Phase diagram and thermodynamics}

We will consider the ground state of our system as a function of the parameters and external sources of the theory. Parameters of the theory are the exponents $\nu,\tau,\gamma$ and the conformal dimension (bulk mass) $\Delta_\chi$. The ranges of the allowed values of $\nu,\tau,\gamma$ are chosen in such a way that the dependence on their values is smooth and unlikely to lead to phase transitions; furthermore, these exponents characterize the running couplings in field theory, which include also the information at different energy scales and probably cannot be realistically tuned. Therefore, the dependence of the thermodynamic quantities on $\nu,\tau,\gamma$ will not be explored. The typical procedure in holographic superconductor literature would be to tune $\Delta_\chi=D/2+\sqrt{D^2/4+m_\chi^2}$ as a proxy for coupling strength in field theory, and this is what we shall do. The requirement for condensation fully fixes the solution $\chi(z)$, as we remind below, and we have no sources for $O_\chi$. However, there is one free parameter in the theory at fixed parameter values: the operator $O_\Phi$ dual to the dilaton in the UV. Therefore, the phase transitions are driven by dialing the scaling dimension $\Delta_\chi$ and the expectation value of the operator $O_\Phi$ dual to the dilaton. When not explicitly stated, we will assume a fixed $O_\Phi$ and study the phase transitions as a function of $\Delta_\chi$.

\subsection{The condensation of the boson at $T=0$}

We expect that at some $\Delta_\chi=\Delta_c$ the neutral bosonic operator $O_\chi$ acquires a nonzero expectation value. As we know \cite{hhh}, the expectation value in field theory is given by the subleading term in the UV expansion (\ref{chibnd}) at zero source term
\be
\label{oeq}\langle O_\chi\rangle=\chi_+\vert_{\chi_-=0}.
\ee
One can also consider an alternative quantization where the VEV is given by $\chi_-$, provided both terms are normalizable, but we will stick with the standard quantization. At this place one should differentiate between the neutral and the charged case. In the neutral case, no continuous symmetry is broken and the phase transition is more akin to nucleation, where the oscillating modes of the scalar add up to a significant perturbation which eventually changes the metric. Whether the oscillations are strong enough or not to lead to a new ground state in principle depends on the parameters of the system. The charged case is expected to be simpler: here, the instability is supposed to be rooted in the Higgs mechanism which breaks the $U(1)$ symmetry, and one expects this to happen for any charged scalar (independently of the $m_\chi^2$ value). The charged scalar is thus expected to always condense at $T=0$, at least in absence of the dilaton. In the presence of the dilaton, things can become more complicated, as we shall see.

\subsubsection{The neutral case}

The critical value of the conformal dimension\footnote{We will use the conformal dimension $\Delta_\chi$ and the bulk mass squared $m_\chi^2$ interchangeably as they are uniquely related to each other through $\Delta_\chi=D/2+\sqrt{D^2/4+m_\chi^2}$.} $\Delta_c$ can be related to the violation of the Breitenlohner-Freedman (BF) stability bound in the interior. To remind, the idea is to rewrite the Klein-Gordon equation for the scalar with energy $\omega$ as an effective Schr\"odinger equation for the rescaled scalar $\tilde{\chi}(z)=\chi(z)/B(z)$ with energy $\omega^2$:
\be
\label{chischeq}\tilde{\chi}''-V_\mathrm{eff}(z)\tilde{\chi}=-\frac{\omega^2}{f^2}\tilde{\chi}(z)
\ee
and the effective potential
\be
\label{chischveff}V_\mathrm{eff}=\frac{e^{-2A}}{f}m_\chi^2-\frac{B'}{B}\left(\frac{f'}{f}+\frac{\partial_\Phi Z}{Z}\Phi'-(D-1)A'\right)-\frac{B''}{B},
\ee
where the rescaling factor is
\be
\label{chischb}B(z)=\frac{e^{-\frac{D-1}{2}A}-\frac{\partial_\Phi Z}{2Z}\Phi}{\sqrt{f}}.
\ee
If the energy of $\chi$ becomes imaginary, i.e. the Schr\"odinger energy $\omega^2$ becomes negative, it means there is an exponentially growing mode which likely signifies an instability, and the scaling dimension becomes complex \cite{BFbound}. In the Schr\"odinger formalism, it means that $\tilde{\chi}$ forms a bound state. We are not allowed to violate the bound in the UV, to prevent violating the AdS asymptotics assumed in the gauge/gravity duality, but an instability in the interior is perfectly allowed and signifies the change of IR physics, i.e. of the field theory ground state. In AdS-RN background, the instability of the neutral scalar is given simply by the BF bound of the near-horizon AdS${}_2$, which equals $-1/4$ \cite{hhh,holscneutral}. We do not have a near-horizon AdS region and there is no simple formula for the critical mass (dimension) $m^2_c$ ($\Delta_c$) but the logic is the same: we are looking for complex energies, i.e. bound states in the Schr\"odinger formalism.

In geometry Ia the effective potential reads
\be
\label{Veffia}V_\mathrm{eff}=m_\chi^2e^{-2z^\nu}-\frac{(D-\gamma-1)\nu(\nu-1)}{2}z^{\nu-2}+\frac{(D-\gamma-1)^2\nu^2}{4}z^{2\nu-2}
\ee
which is positive and growing to infinity at large $z$. For any bound states to exist, we need to have a sufficiently deep and broad potential well below zero energy, i.e. the potential needs to grow to infinity also on the "left-hand side", for small $z$, and fall sufficiently low in-between. Now we remember that for $z\to 0$ the potential certainly goes to positive infinity because $m_\chi^2>-D^2/4$ (i.e., we do not want bound states in the far UV region, sitting at $z\to 0$). Now the question is what the potential looks like for some intermediate $z_1$ which is still large enough that the IR solution (geometry Ia) is valid. Assuming that $z_1\sim 1$, this depends on the combination $m_\chi^2-(D-\gamma-1)\nu(\nu-1)/2+(D-\gamma-1)^2\nu^2/4$ -- the second and third term are both positive, and the question is whether there is a value of $m_\chi^2>-D^2/4$ which is nevertheless sufficiently negative to make $V_\mathrm{eff}$ negative. This is obviously a question of numerical calculation but we can see that for $\gamma=D-1+\epsilon$ for $\epsilon$ small the second and the third term in (\ref{Veffia}) have practically zero coefficients and not too large $\vert m_\chi^2\vert$ suffices to push $V_\mathrm{eff}$ below zero in some interval. We conclude that we can expect a BF-type instability at some critical $m_c^2$. We have seen this means the geometry Ia is modified, presumably into IbC, and at finite $m_c^2$, analogously to the neutral holographic superconductor in AdS-RN \cite{hhh,holscneutral}.

\begin{figure}[ht!]
\begin{center}
(A)\includegraphics[width=0.45\linewidth]{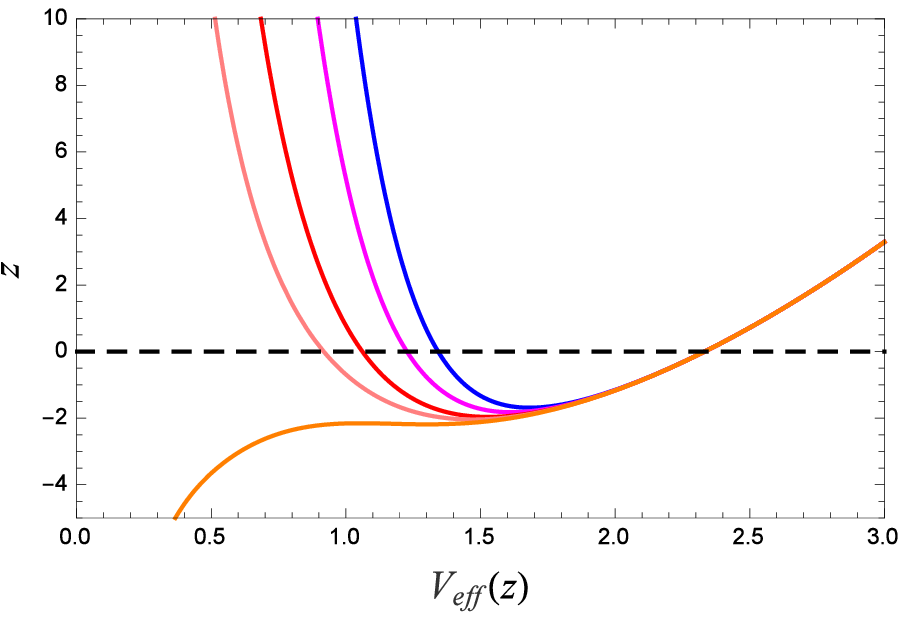}
(B)\includegraphics[width=0.45\linewidth]{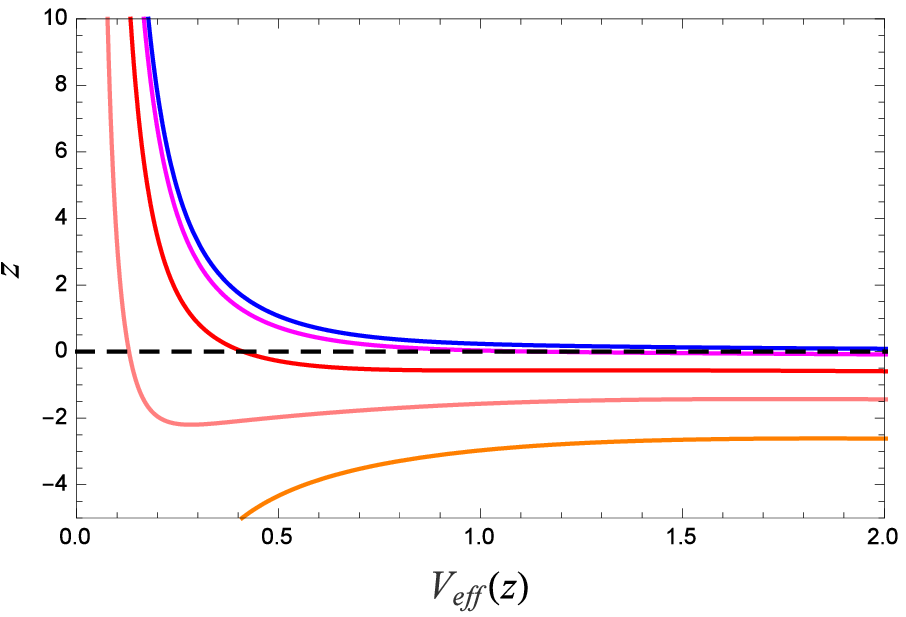}
\caption{The effective Schr\"odinger potential $V_\mathrm{eff}(z)$ defined by (\ref{chischveff}) for a range bulk masses (conformal dimensions) $m_\chi^2=6,2,0,-2,-4$ (blue, magenta, red, pink, orange) and $D=4,\nu=2,\gamma=4$. The instability corresponds to bound states, i.e. existence of a sufficiently deep and broad potential well. In (A), we can fit a bound state for all masses shown, for the last one just a single bound state, thus $m_c^2\sim 6$ corresponds to the BF bound. For such masses, the geometry will remorph and we will enter the condensed phase. This phase is stable, as in (B) the potential well is too shallow to accommodate a bound state. Notice that for $m_\chi^2=-4$ the potential develops a well in the outer region, i.e. this is the BF bound for AdS${}_5$.}
\label{figpotneut}
\end{center}
\end{figure}

Having shown that there is indeed a mechanism for the condensation of the order parameter in the soft-wall regime, we should also check if the geometry IbC is stable in the presence of the condensate. In geometry IbC the effective potential is:
\be
\label{Veffib}V_\mathrm{eff}=V_\infty+\frac{m_\chi^2}{z^{2\alpha}}+\frac{\kappa(\kappa+1)}{z^2}
\ee
where $\kappa=\frac{(D-1)\alpha+\gamma\phi_0}{2}$ and $V_\infty$ is a $z$-independent constant. The inverse square term is always positive and the power of the mass term varies between $-\infty$ for $\gamma\to D-1$ and $-2$ for $\gamma=2D$ (we see this from the expressions for $\alpha,\phi_0$ in (\ref{geoIbC})). Thus the $1/z^2$ term dominates at large $z$ for the allowed values of $\gamma$ (from (\ref{dilpotz})) and approaches zero from above as $z\to\infty$; this means the potential approaches the constant $V_\infty$ from above. This in turn means there is no room for bound states -- the potential in the UV is positive and decaying and never falls below zero.\footnote{This picture changes for $\gamma>2D$ -- then the mass term dominates for $z\to\infty$ and for negative mass squared it forms a potential well. But in our model one always has $\gamma<2D$ so we do not explore this case in detail.} Therefore, the geometry IbC is stable in the presence of the scalar. Numerical plot of the potential in Fig.~\ref{figpotneut} confirms the above discussion. In the panel (A) there is a potential well with bound states for all masses below some $m_c^2\sim 6$ which is thus the critical value for the condensation. In panel (B) the well turns out too shallow to allow the formation of bound states: the geometry is stable. All curves are for $m_\chi^2\ge -D^2/4$ as for this value there is a potential well near $z=0$ and the outer AdS region becomes unstable.

\begin{figure}[ht!]
\begin{center}
(A)\includegraphics[width=0.45\linewidth]{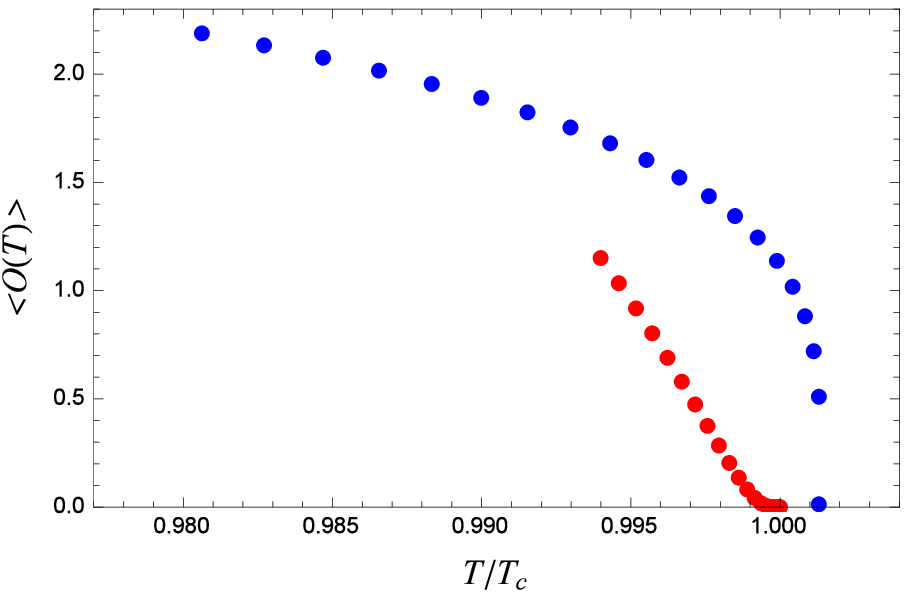}
(B)\includegraphics[width=0.45\linewidth]{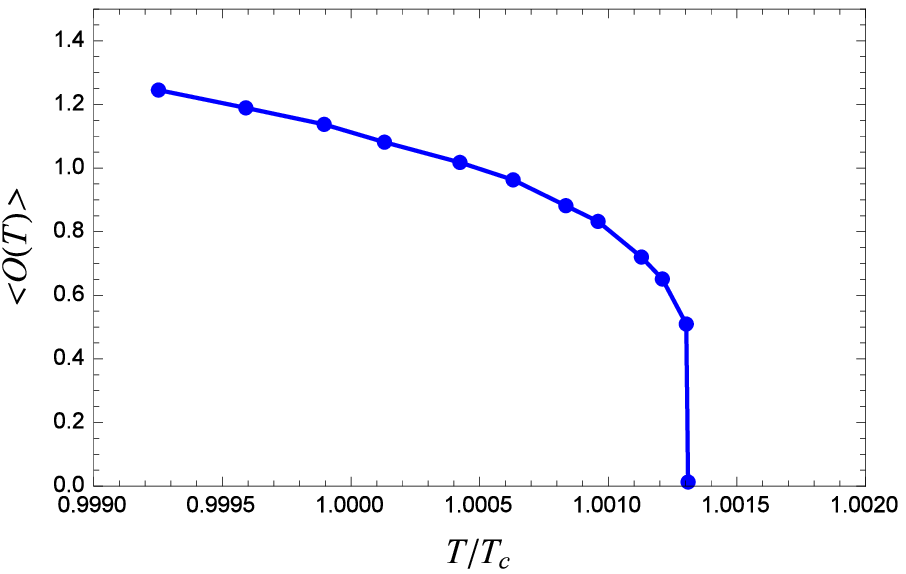}
\caption{(A) Expectation value of the scalar $\langle O_\chi\rangle$ as a function of temperature for $m_\chi^2=-2$, for the neutral scalar (blue) and the charged scalar with $q=1$ (red), in $D=4$ and for $\nu=2,\gamma=4$; for the charged scalar $\tau=5$. The neutral scalar has a first-order quantum phase transition and its value jumps from zero, whereas in the charged case the quantum phase transition shows a continuous BKT-like exponential form $\exp\left(-\left(T_c-T\right)^{-1/2}\right)$. The unit of temperature is $T_c$ -- the critical temperature for the \emph{charged} case. In  (B) we zoom-in near the critical temperature for the neutral case to make it obvious that there is a jump.}
\label{figcond}
\end{center}
\end{figure}

In the numerical calculation, we shoot for the solution of a two-point boundary value problem which satisfies the boundary condition (\ref{chibnd}) for $\chi$ at the AdS boundary and the expected asymptotics for $\chi(z)$ from (\ref{geoIbC}) in the interior. We do this as a part of the complete calculation (with backreaction on geometry, see the Appendix). In this way we can find the dependence of the VEV $\langle O_\chi\rangle$ on the conformal dimension $\Delta_\chi$. In Fig.~\ref{figcond}(A), the blue curve jumps at the transition, signifying that the transition is of first order. This is different from the infinite-order BKT-type (stretched-exponential) scaling laws found in \cite{hhh,holscneutral} for a neutral scalar in AdS-RN background because the BKT scaling originates in so-called Efimov states in the IR which depend on the details of the potential for the scalar \cite{holscneutral} and would require a fine tuning of the dilaton potentials too. First-order transition is not unknown even for a charged scalar if it is non-minimally coupled to the metric \cite{yanyawendil}. We also expect the condensate to vanish at higher temperatures, a case which we find too difficult for analytical work so we limit ourselves to numerics. The result is shown in Fig.~\ref{figcond}: there is again a jump at the critical temperature.

\subsubsection{The charged case}

The charged problem can usually be understood as the textbook Abelian-Higgs instability where the gauge field develops an effective mass term $\vert\chi\vert^2A_0$ and the mass of the scalar is effectively negative as it acquires a correction $-g^{00}A_0^2$, leading to instability and condensation. Without dilaton, in AdS-Reissner-Nordstrom background, this correction to the scalar mass grows fast enough near the horizon to produce an instability even at positive $m_\chi^2$ \cite{horroberts}. For our system the equation for the charged scalar in IR geometry IaQ (Eq.~\ref{geoIaQ}) reads
\be
\label{a0eqiaq}\chi''-\nu(\tau+3-\gamma-D)z^{\nu-1}\chi'-\frac{e^{(\tau-2D+2)z^\nu}}{\xi}\left(m_\chi^2-\xi a_1^2q^2z^{2\nu-2}e^{(\gamma-\tau)z^\nu}\right)\chi=0
\ee
Now the negative correction to the effective mass of the scalar may grow or diminish as $z\to\infty$, depending essentially on the sign of $\gamma-\tau$. If $\gamma>\tau$ the correction dominates the bare mass term and we always have a mode growing at $z\to\infty$ but if $\gamma<\tau$ it is subleading and does not influence the behavior of $\chi(z\to\infty)$ at leading order. Looking at our conditions (\ref{dilpotv}-\ref{dilpotz}), we see this is always the case. Naively, one may guess that the critical value is $m_c^2=0$ but since our analysis ignores all subleading terms one should check numerically (numerics confirms that this is indeed the critical value, see the phase diagram in Fig.~\ref{figphdiag}). Amusingly, the scaling with temperature and conformal dimension is now consistent with the BKT-like form:
\be
\label{chgscaling}\langle O_\chi\rangle=\mathrm{const.}\times e^{-\frac{1}{\sqrt{-m_\chi^2}}}.
\ee
Although the numerical fit to the $e^{-1/(-m_\chi^2)^n}$ law with $n=1/2$ is good, we cannot exclude the possibility that the exponent $n$ weakly depends on $\nu$ and that it is not exactly $1/2$; we have no analytical estimate for $n$. The condensate formation is now, strictly speaking, not a consequence of the coupling with the gauge field at all (remember the term $q^2g^{00}A_0^2$ is now exponentially suppressed) but merely the consequence of growing modes for negative scalar mass. Thus the mechanism is essentially the same as for the neutral scalar and the fact that the neutral scalar undergoes a discontinuous transition reminds us that the details of this process depend sensitively on the IR geometry. The temperature scaling is of the same form as the scaling with $m_\chi^2$ (\ref{chgscaling}) and is shown as red points in Fig.~\ref{figcond}.

\subsection{Free energies and phases at zero temperature}

Now that we have explained the instability that seeds the condensation, we will compute the free energy (on-shell action) of the system as a function of $\Delta_\chi$ and $T$, to study the order of the transition and the full phase diagram. We thus need to evaluate (\ref{action0}) on-shell for solutions Ia and IbC: $\mathcal{F}=\int d^DxL\vert_{Ia,IbC}+\mathcal{F}_{bnd}$. The boundary terms are given by
\be
\label{frenbnd}\mathcal{F}_{bnd}=\oint_{bnd}\sqrt{g_\mathrm{ind}}(-2K-\lambda-\frac{1}{2}A_0A_0'-\chi^2-2\Phi\Phi'),
\ee
Here, $g_\mathrm{ind}$ is the induced metric at the boundary, $K$ is the trace of the extrinsic curvature, $\lambda$ is the boundary cosmological constant, and the remaining terms come from the gauge field, the scalar and the dilaton. The counterterm for the scalar is in accordance with our choice that $\chi_+$ is the VEV; had we chosen $\chi_-$ for the VEV the counterterm would be $-2\chi'\chi$, analogous to the situation for the dilaton. The comparison of free energies is best done numerically but even analytically we can draw some conclusions. Let us first consider the quantum phase transitions as a function of $\Delta_\chi$ at fixed $\Phi_-$ and discuss the free energies at zero temperature.\footnote{At $T=0$ the free energy is just the total energy $\mathcal{E}$ of the system, since $\mathcal{F}=\mathcal{E}-TS$. For simplicity of notation, we will still call it $\mathcal{F}$ just like the finite temperature case.} Our analytical solutions are only valid in the large $z$ region, whereas for smaller $z$ they cross over into the AdS${}_{D+1}$ forms, so the radial integral in (\ref{frenbnd}) goes from some $z_1\sim 1$. The difference between the energy of the solution Ia (we will show numerically it is indeed preferred to Ib) and IbC is
\be
\label{frenibQ}\mathcal{F}_{Ia}-\mathcal{F}_{IbC}\sim\chi_+^2z^{2\Delta_\chi}+\int_{z_1}^\infty dz\left[(D^2-D)z^{-(D-1)\alpha}+\chi_0^2m_\chi^2z^{-\gamma\phi_0}+\ldots\right].
\ee
The difference in free energies at leading order has terms proportional to the squared amplitude of the order parameter (in the UV -- $\chi_-$, i.e. $\langle O_\chi\rangle$ and in the IR -- $\chi_0$) but also a $\chi$-independent term (coming from the Ricci scalar and cosmological constant terms in geometry IbC) so we expect that the transition, determined by $\mathcal{F}_{Ia}-\mathcal{F}_{IbC}=0$ generically happens at nonzero amplitudes $\langle O_\chi\rangle,\chi_0$ and we can exclude a continuous transition. This is again in line with the discreteness of the symmetry broken and the discontinuous nature of the transition. On the other hand, for the charged geometries IaQ and IbQC there is also the boundary contribution $A_0(z\to 0)A_0'(z\to 0)$ so
\be
\label{freniaQ}\mathcal{F}_{IaQ}-\mathcal{F}_{IbQC}=\frac{\mu(\rho_{IaQ}-\rho_{IbQC})}{2}+\chi_+^2z^{2\Delta_\chi}-\int_{z_1}^\infty dzz^{-(D-1)\alpha}\chi_0^2(4+7\gamma)^2+\ldots
\ee
Now there is no $\chi$-independent term and the dominant terms in the energy difference are proportional to the squared amplitude of the condensate, or to the difference in charge densities $\rho_{IaQ}-\rho_{IbQC}$ which, according to the Gauss-Ostrogradsky theorem, also has to be proportional to the bulk density of the charged field, $q^2\chi(z)^2$. Therefore, one can expect that the energy difference grows from zero at $\langle O_\chi\rangle=0$, as in a continuous phase transition. We have assumed that the chemical potential is kept constant across the transition (grand canonical ensemble). Now we will check our conclusions numerically.

\begin{figure}[ht!]
\begin{center}
(A)\includegraphics[width=0.45\linewidth]{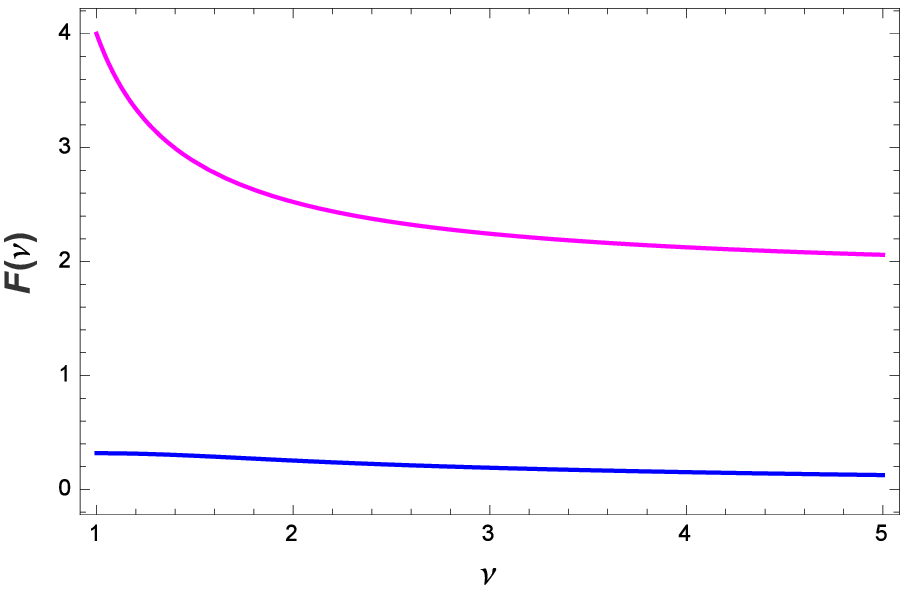}
(B)\includegraphics[width=0.45\linewidth]{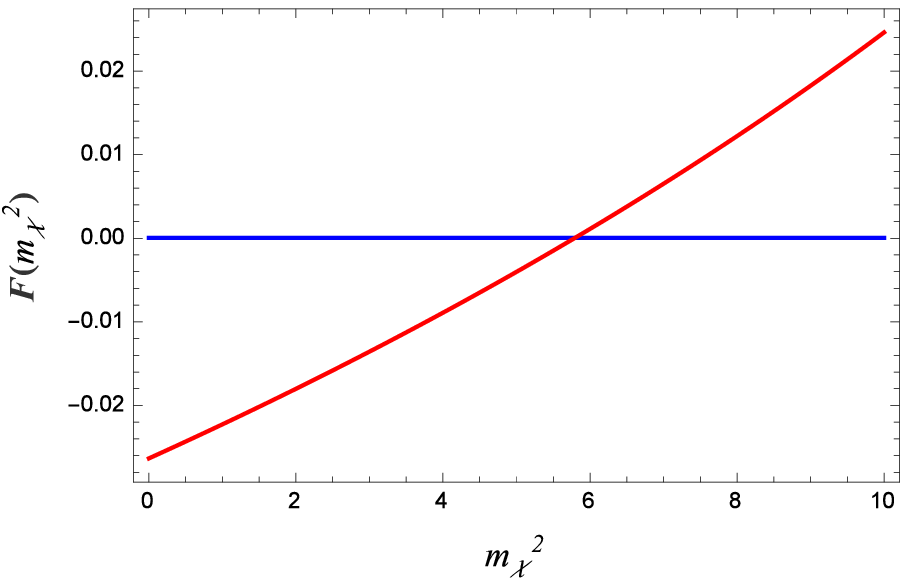}
\caption{(A) Free energy at zero temperature as a function of the scaling exponent $\nu$ in the absence of condensate, for geometry Ia (confining, blue)and Ib (nonconfining, magenta). Obviously the soft-wall geometry always has lower free energy, thus it is always preferred: in absence of condensate we have a confining soft wall. The units on the vertical axis are arbitrary. (B) Free energy at zero temperature as a function of the bulk mass $m_\chi^2$ for geometry Ia (confining, no condensate, blue) and for geometry IbC (nonconfining, with condensate, red). Both solutions exist before and after the critical point, where their energies $F(m_\chi^2)$ intersect at finite angle, thus the phase transition is of first order. The solid, dashed and dotted lines are from three different values of the source $\Phi_-=0.1,0.2,0.5$ -- the source shifts the location of the transition but does not change the behavior qualitatively. The free energy is in computational units and the parameters are $\nu=3/2,\gamma=D=4$.}
\label{figfrenab}
\end{center}
\end{figure}

First of all let us show that the confined solution is indeed the ground state in absence of the condensate. In Fig.~\ref{figfrenab}(A) we plot the on-shell action of the solutions Ia (\ref{geoIa}) and Ib (\ref{geoIb}) and we see that Ia indeed always has lower energy -- the system is confining. Now consider the free energies as functions of $m_\chi^2$ and the temperature. In Fig.~\ref{figfrenab}(B) we compare the free energies as functions of the conformal dimension for the neutral system and confirm the discontinuous nature of the transition: the curves have different derivatives at the transition point. Here we also scan for different values of the source $\Phi_-$, which change the value of the transition point $\Delta_\chi$ but, importantly, do not introduce new phases. This is easily understood from the discussion in section IV.A.1 and also from Eq.~(\ref{frenibQ}). Dialing $\Phi_-$ influences the matching between the solutions in the UV and the solutions in IR without introducing new IR solutions, so we are still left with the choice between Ia and IbC. Concerniing the scalar condensation, different values of $\Phi_-$ reshape the effective potential, influencing the point $z_1$ where the geometry crosses over to the IR asymptotics and thus the width of the potential well, so it starts supporting bound states for different values of $m_\chi^2$. Finally, the free energy difference depends on the IR quantities $\phi_0,\chi_0$ which are determined by the matching to the UV solution. Their values influence the location of the transition point but not the nature of the transition.

\begin{figure}[ht!]
\begin{center}
(A)\includegraphics[width=0.45\linewidth]{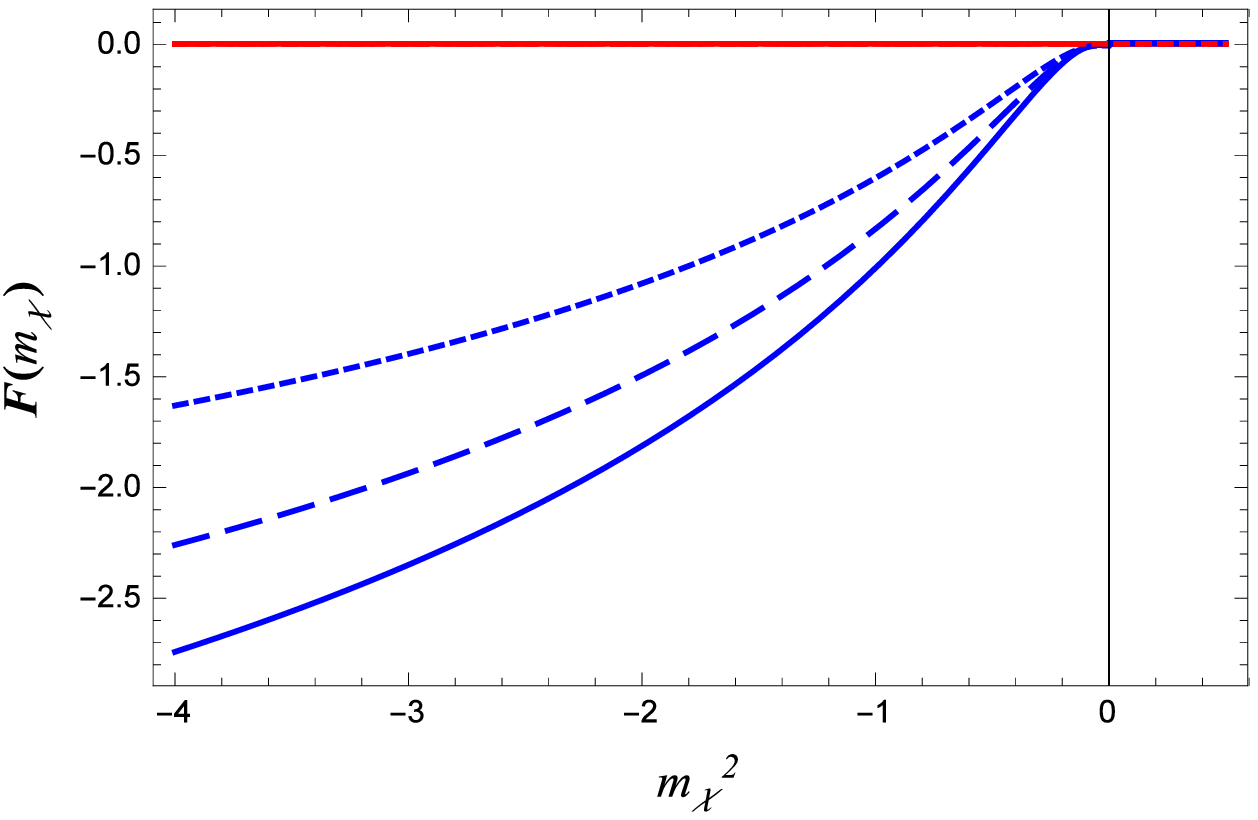}
(B)\includegraphics[width=0.45\linewidth]{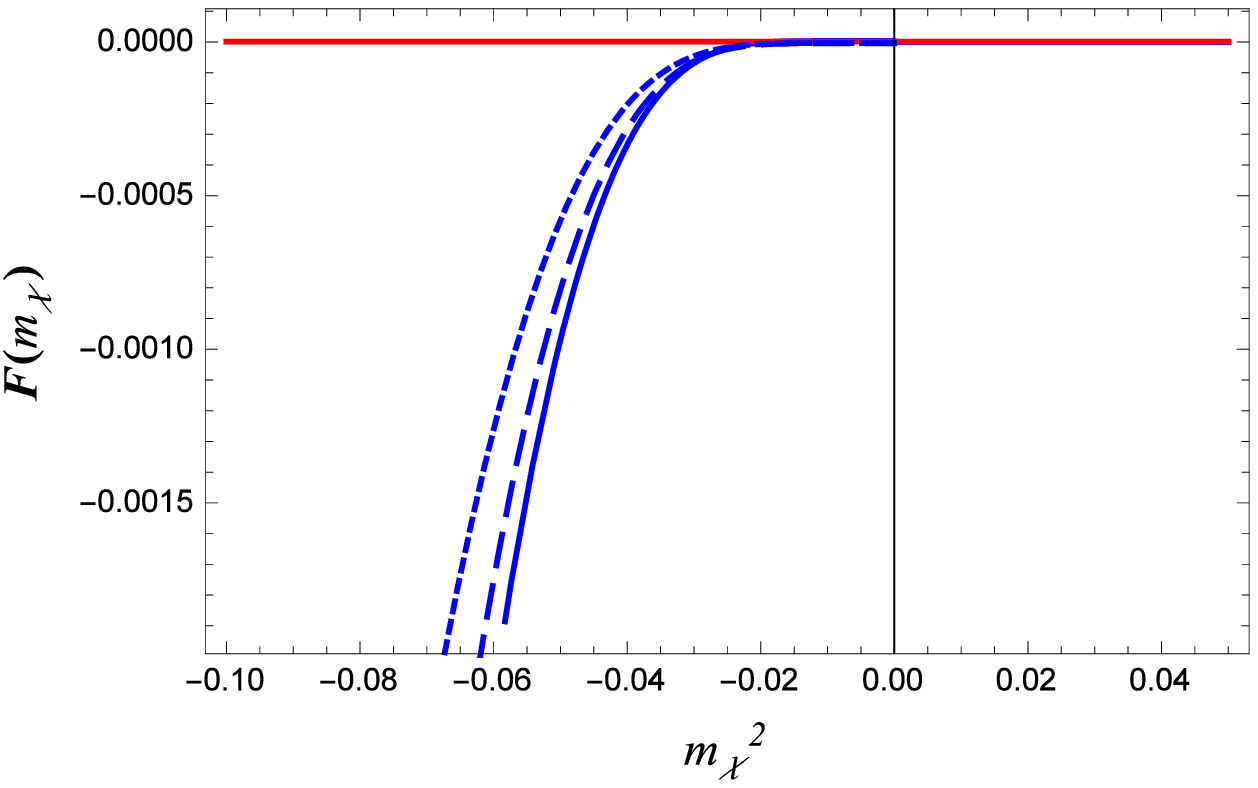}
\caption{Same as in Fig.~\ref{figfrenab}(B) but for the charged order parameter with $q=1$ (A), with a zoom-in near $m_c^2=0$ (B). The new solution with condensate branches off smoothly and with continuous first derivative, thus the phase transition is continuous. It is consistent with BKT-like scaling $e^{-1/\sqrt{\Delta_c^2-\Delta^2}}$. The three values of source (solid, dashed, dotted lines, same as in the previous figure) leave the critical point $m_c^2=0$ invariant and only influence the deconfined, condensed phase. The free energy is in arbitrary units and $\tau=6$.}
\label{figfrenchg}
\end{center}
\end{figure}

For the charged case the free energy is given in Fig.~\ref{figfrenchg}. The transition is now continuous and the zoom-in near the origin clarifies that the critical point lies at $m_\chi^2=0$. Interestingly, the three values of the $O_\Phi$ source now all give the same critical point, at zero mass squared. The curves for different values of $O_\Phi$ only differ in the deconfined phase, with nonzero $\langle O_\chi\rangle$, and coincide as long as no condensate forms. At first, this may sound strange. However, a look at the effective potential (\ref{a0eqiaq}) shows that the negative term is now exponentially growing at large $z$ and thus the potential well is always in the deep IR region, rather than in the middle as in the neutral case (Fig.~\ref{figpotneut}). It is thus understandable that it is not affected by the matching to the UV solution with given $\Phi_-$.

We have already established that our confinement/deconfinement transition may be of continuous or discontinuous nature. Both cases are in principle known even in field theory, and all the more so among the many condensed matter systems where some kind of fractionalization picture is appropriate.

\subsection{Finite temperature thermodynamics}

At finite temperature, the free energy is still the value of the on-shell action but the radial integration now terminates at finite $z_h$. In the leading term of the action in geometry Ia, which stems from the dilaton potential:
\be
\label{frenia}\mathcal{F}_{Ia}\sim -V_0\int_{z_1}^{z_h} dz\frac{e^{-(D-3)z^\nu}}{z^2}+\ldots
\ee
we need to perform the radial integration from the crossover-to-AdS-scale $z_1$ to the horizon $z_h$ and expand the result about $z_h$. Therefore, we integrate from "deep IR" at $z_h$ to "the UV of the IR", i.e. the location where the geometry crosses over to the asymptotic AdS${}_{D+1}$. Clearly, the integral is dominated by the exponential term and our free energy scales as
\be
\label{frtia}\mathcal{F}(T\to 0)\propto\frac{1}{z_h^2}\Gamma_{2-1/\nu}((D-3)z_h^\nu)\sim\mathrm{const.}\times e^{-\frac{D-2}{T^\nu}}T^{3-\nu}
\ee
where the power-law correction $T^{3-\nu}$ is in fact unimportant (we don't consider $D<3$ so the exponent $-(D-2)/T^\nu$ is always negative) and the free energy has an extremely slow growth at low $T$. Clearly, the entropy $S=\partial\mathcal{F}/\partial T$ is zero at zero temperature, and is extremely low at low $T$ (much smaller than for any system with the scaling $\mathcal{F}\sim T^x$ for any power $x$). Thus the effective number of the degrees of freedom is much reduced because of the confinement. The same scaling is obtained for the charged case.\footnote{One may wonder whether this slow growth of entropy can actually be observed. It is possible that any amount of disorder in the system would make the entropy significantly larger. At least theoretically, however, an exponentially slow growth is not unusual in dilatonic setups, see e.g. \cite{umut}.} At high temperatures (compared to the confinement gap) we can expand the action in $1/T$ and get
\be
\label{frtiahigh}\mathcal{F}(1/T\to 0)\propto\frac{\Gamma(1-1/\nu)}{z_h^2}\sim\mathrm{const.}\times T^2,
\ee
the quadratic behavior of the free energy and the linear behavior of entropy characteristic of Fermi liquids.\footnote{For high temperature we get $T\sim 4\pi D/z_h$ but this is not en exact relation and is not even close at low $T$ (unlike the textbook Schwarzschild or RN black hole without the dilaton).} This result was found for a dilatonic black hole in \cite{gubser1} and our system behaves similarly at high temperatures (in fact, our confined charged system only differs from it by the choice of the dilaton potential, which likely influences the low-temperature behavior but not the high-$T$ asymptotics). Even though we have no fermions in the system, the quadratic scaling is perhaps not so surprising: one may expect it in any confined system, where only the gauge-neutral bound states are observable. Notice, however, that at high temperatures we expect a dimensional scaling to take place

\begin{figure}[ht!]
\begin{center}
(A)\includegraphics[width=0.45\linewidth]{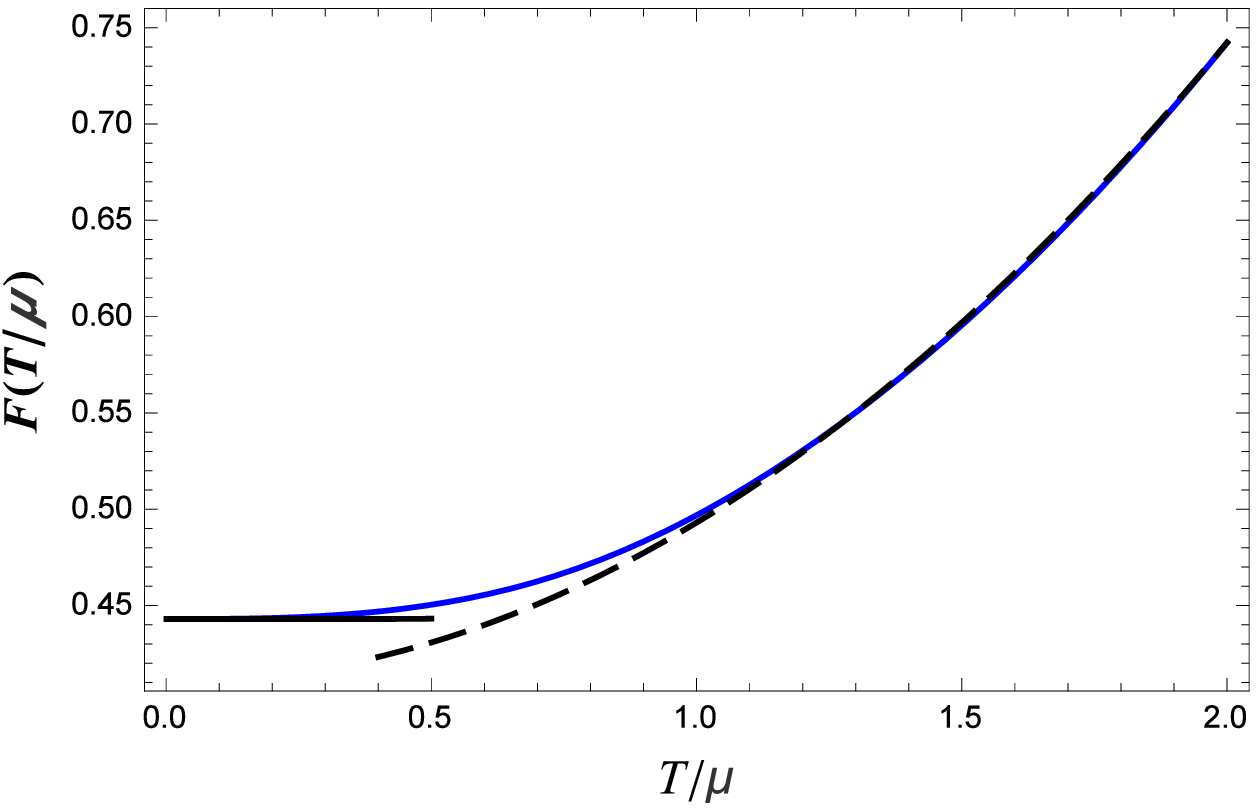}
(B)\includegraphics[width=0.45\linewidth]{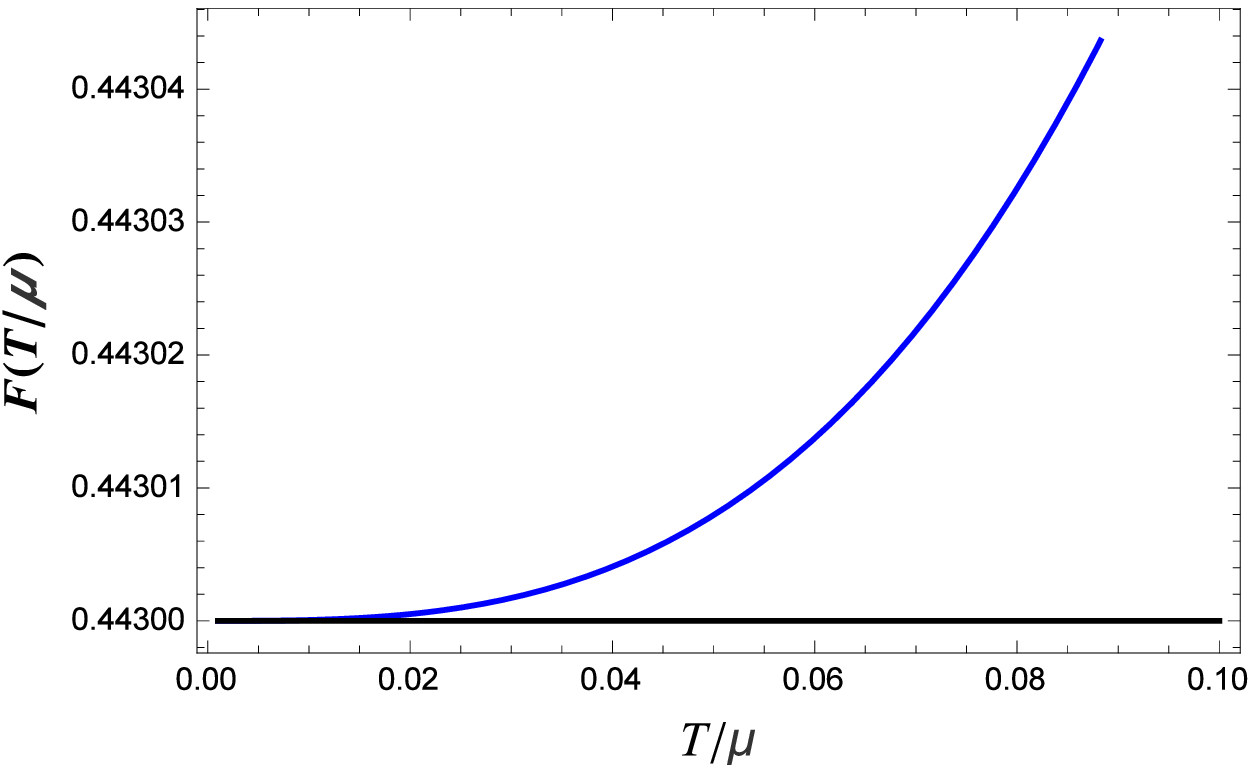}
\caption{Temperature dependence of the free energy $\mathcal{F}(T)$ for the charged system in the confined phase ($m_\chi^2=1$), for $\mu=1,\nu=2,\gamma=4,\tau=6$. The dashed and the dotted black lines are the analytical estimates (\ref{frtia}, \ref{frtiahigh}) for the confined phase. In (A) we cover a broad range of temperatures, showing both the low-temperature regime with the scaling (\ref{frtia}) and the high-temperature regime (\ref{frtiahigh}). In (B), we zoom in at low temperatures, showing the very slow growth of free energy and entropy. The analytical estimates for the low-temperature scalings are determined only up to the UV contribution, which was assumed approximately constant and was fit to the numerics.}
\label{figfrt}
\end{center}
\end{figure}

In the deconfined phase, the exponential scaling is gone and we have a simple scaling law for both low and high temperatures:
\be
\label{frtib}\mathcal{F}(T)\propto z_h^{-(D-1)\alpha}\sim\mathrm{const.}\times T^x.
\ee
For high temperatures the exponent is $x=(D-1)\alpha$; for low $T$ the relation $T(z_h)$ is complicated but behaves as a power-law, so $\mathcal{F}$ still scales as a power law of the temperature. This anomalous power law for all temperatures is precisely in line with the hyperscaling-violating nature of the system: the metric has power-law scaling and has no sharp scale where low-$T$ regime cross over to high-$T$ regime. These findings are illustrated in Fig.~\ref{figfrt}, where we plot the numerical calculation of $\mathcal{F}(T)$ together with analytical scaling laws for the confined system. We have chosen a large and positive scalar mass $m_\chi^2=4$ to avoid the phase transition to the condensed deconfined system, since the purpose of the figure is to study the different scaling regimes in the same phase, not the phase transition (which is discontinuous in $T$ for the neutral system and of infinite order in $T$ in the charged system, same as the scaling with $m_\chi^2$). Notice that the analytical estimates (\ref{frtia}-\ref{frtib}) are only the IR contribution, and the true free energy is obtained by adding the UV contribution, which is fit as a constant in Fig.~\ref{frt} (assuming that the $T$-dependence in the UV is weak, though in reality it is certainly not strictly constant).

\subsection{Structure of the phase diagram}

\begin{figure}[ht!]
\begin{center}
\includegraphics[width=0.90\linewidth]{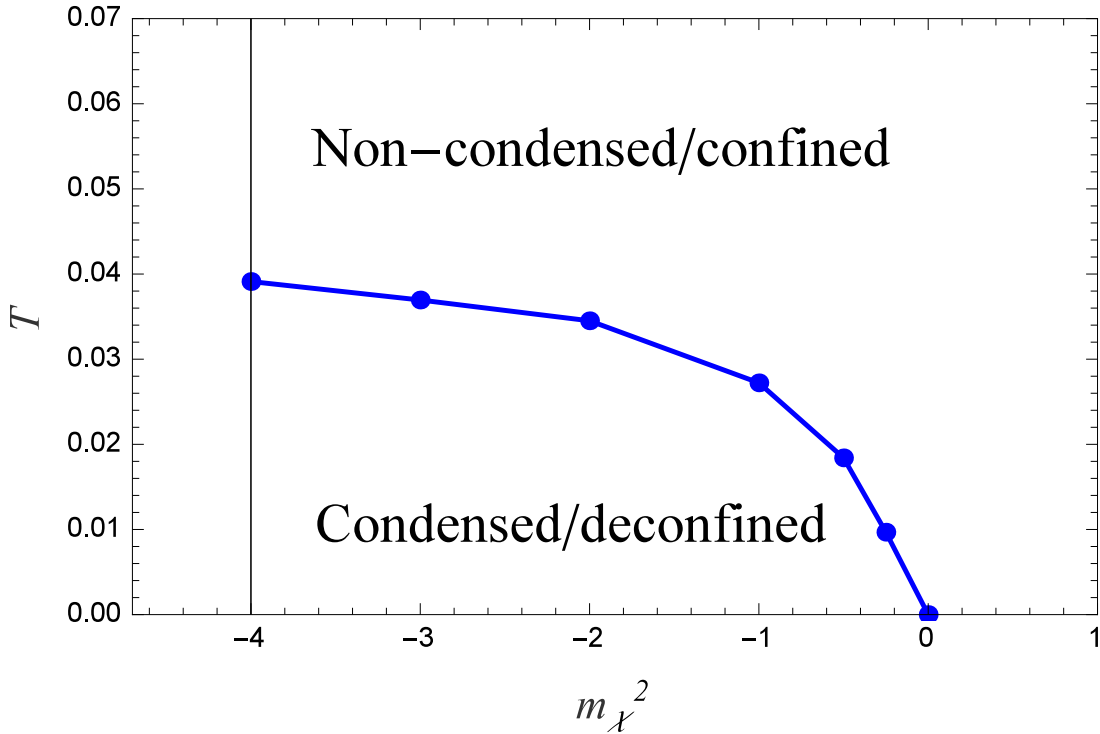}
\caption{Phase diagram in the $\Delta_\chi$--$T$ plane for the charged scalar. Blue dots denote the numerical results for the onset of the condensation of the scalar; the line is just to guide the eye. The condensed/deconfined phase (geometry IaQ) is located to the left and below the boundary line; the rest is the non-condensed/deconfined phase (geometry IbQC). For the neutral case the phase diagram is similar. The key finding is that the deconfinement transition coincides with the onset of the long-range order. The vertical black line denotes the BF bound for AdS${}_5$.}
\label{figphdiag}
\end{center}
\end{figure}

We are now in position to construct the whole phase diagram. The phases are the same both for the neutral and for the charged case, except that the critical line is of different nature (first-order and smooth, respectively). The phase diagram is sketched in Fig.~\ref{figphdiag}. For small enough conformal dimension $\Delta_\chi$ and temperature $T$, the scalar condenses and the system deconfines, restoring the scale invariance at low energies. As the temperature rises, the long-range order of the scalar is lost and we are back to the confined regime. This shows our main point -- the confinement/deconfinement transition is triggered by the long-range order of $O_\chi$. What does this mean symmetry-wise? On one hand, the condensation of $O_\chi$ certainly breaks a symmetry -- $\mathbb{Z}_2$ (neutral) or $U(1)$ (charged). But on the other hand the deconfinement restores a symmetry: as we have explained, the deconfined geometries are anisotropically scale-covariant (hyperscaling), of the form $ds^2=z^{-2\kappa}(-f_0z^{-\eta}dt^2+d\mathbf{x}^2+f_0^{-1}z^\eta dz^2)$. In absence of charge ($f_0=1,\eta=0$), all coordinates in field theory can be rescaled as $x^\mu\mapsto\lambda x^\mu$ though the energy (dual to $z$) scales differently (this is sometimes called generalized conformal symmetry, \cite{bigkir}). With nontrivial $f$ the scaling exponent is different along different axes but there is still some invariance to dilatatons (rescaling of coordinates). At the same time, in the soft-wall case with $ds^2\propto e^{-2z^\nu}$ there is no scale invariance at all. Overall, neither phase is more symmetric than the other: denoting the symmetry group of the scaling system in field theory by $\mathbb{G}_1$, we expect it to be broken in the confined phase down to some subgroup $\mathbb{G}_2<\mathbb{G}_1$, while the symmetry of the scalar ($\mathbb{Z}_2$ or $U(1)$) is fully broken in the deconfined phase. Since we have a bottom-up model we don't have the explicit form of the field theory Lagrangian and so we cannot fully determine $\mathbb{G}_{1,2}$. Both certainly include the spacetime translations and rotations and $\mathbb{G}_1$, as discussed, contains also dilatations. In special cases, e.g. when the field theory is $\mathcal{N}=4$ super-Yang-Mills, it will be the full conformal group and the deconfinement will be the restoration of the full conformal symmetry. In any case, the symmetry at the critical point changes like
\be
\label{symm}\mathbb{G}_2\otimes\mathbb{Z}_2\mapsto\mathbb{G}_1\otimes\mathbb{I},~~\mathbb{G}_2\otimes U(1)\mapsto\mathbb{G}_1\otimes\mathbb{I}.
\ee
The neutral case where the phase transition is discontinuous could be related to the Landau-Ginzburg theory which generically predicts that in such situations, when no overall symmetry reduction occurs, the two phases can be separated by a first-order transition or by a finite area of phase coexistence. But the charged case where the transition is continuous is of non-Landau-Ginzburg type. This case in particular resembles the concept of deconfined criticality proposed as an explanation for the physics of some strongly coupled quantum critical points in $D=3$ \cite{deconfcrit}. We would like to understand how one could probe such phase diagrams in nature, having in mind the handicap that in a bottom-up gauge/gravity model we do not know the explicit form of the action to directly inspect the symmetries of different phases. We would also like to gain a better knowledge of the confinement/deconfinement transition itself: we cannot directly identify the gauge-charged and gauge-neutral degrees of freedom but we can detect the existence of bound states in the confined phase and explore their dispersion relation, a technique particularly used in AdS/QCD, where the quark confinement is recognized from the linear scaling of bound state masses, $m_n\propto n$ \cite{soft,umutkir1}. We can also look for the signs of symmetry breaking in the response functions. Bound states can be detected in this way too, since they manifest as poles of correlation functions in the imaginary half-plane, separated from the possible quasiparticle peak by a gap (the binding energy).

Finally, one should have in mind that at very high temperatures it is possible that both confined and deconfined solutions (i.e., all the solutions we have considered) give way to the solution with zero dilaton profile, i.e. the system becomes just a (neutral, Schwarzschild or charged, Reissner-Nordstrom) black hole, as pointed out in \cite{albert}. This depends on the parameters of the dilaton potentials; for some values such solutions exist and for some not. We have not checked the existence of this regime explicitly and will not consider it; it is not relevant for the low-temperature and zero-temperature phase transitions we consider here.

\section{Response functions and bound states}

\subsection{Definition and equations of motion}

In this section we will try to understand better the nature of different phases by computing the electric AC conductivity $\sigma(\omega,k=0)$ and charge susceptibility $\xi(\omega,k)$ of our system as well as the retarded propagator $G_R(\omega,k)$ of the order parameter $O_\chi$, in particular by looking at the bottom half of the complex frequency plane where one can find the poles corresponding to the bound states typical of confined systems. In this section we consider the $T=0$ case as we are interesting in the properties of the ground state (and its excitations encoded in the pole structure), not the finite-temperature fluctuations. According to the basic dictionary (e.g.~\cite{janbook}) the conductivity, as the response of the current to the imposed (transverse) electric field, is proportional to the ratio of the source and VEV terms of the fluctuation of the spatial component of the bulk electromagnetic field:
\be
\label{conddef}\delta A_x(z;\omega,k)=\delta A_x^{(0)}+\delta A_x^{(1)}z+\ldots,~~\sigma(\omega,k)=\frac{1}{\imath\omega}\frac{\delta A_x^{(1)}}{\delta A_x^{(0)}}+\frac{1}{\imath\omega}R(\omega,k),
\ee
where $R(\omega,k)$ is the regulator connected to the boundary counterterms in the action. Without entering into detailed discussion, we can quote that in $D=3$ no regulator is needed ($R=1$) whereas in $D=4$ we have $R=k^2-\omega^2$ \cite{horroberts}. Charge susceptibility is the response of the charge density to the applied electric field, and therefore can be computed analogously from the fluctuation of $A_0$:
\be
\label{suscdef}\delta A_0(z;\omega,k)=\delta A_0^{(0)}+\delta A_0^{(1)}z+\ldots,~~\xi(\omega,k)=\frac{\delta A_0^{(1)}}{\delta A_0^{(0)}}=\frac{\delta\rho}{\delta\mu},
\ee
so the susceptibility can be interpreted as the ratio of the charge density fluctuation and the fluctuation in chemical potential. The conductivity mainly makes sense at zero momentum (in the absence of a lattice) whereas susceptibility can also be considered as a function of momentum, to study the spatial modulation of the charge density, as in \cite{donos}. The equations of motion are really the variational equations from the action (\ref{action0}-\ref{action3}) about the equilibrium solutions $A_0(z;\omega,k)$ and $A_x(z;\omega,k)=0$:
\bea
\label{deltax}\delta A''_x-\left((D-3)A'-\frac{\partial_\Phi\mathcal{T}}{\mathcal{T}}\Phi'\right)\delta A'_x-\left(\frac{\omega^2}{f^2}-\frac{k^2}{f}-\frac{2q^2e^{-3A}}{f\mathcal{T}}\chi^2\right)\delta A_x=0\\
\label{deltat}\delta A''_0-\left((D-3)A'-\frac{\partial_\Phi\mathcal{T}}{\mathcal{T}}\Phi'\right)\delta A'_0-\left(\frac{\omega^2}{f^2}-\frac{k^2}{f}-\frac{4q^2e^{-3A}}{f\mathcal{T}}A_0\chi^2\right)\delta A_0=0.
\eea
Even though the fluctuations $\delta A_0, \delta A_x$ are coupled to the fluctuations of the metric, we do not consider the full system of fluctuation equations here. For a charged system, this amounts to working in the limit of large charge, where the probe barely has any influence on the system.

Finally, to study the symmetry breaking we explore also the fluctuation of the scalar field $\delta\chi$ which determines the retarded propagator $G_R(\omega,k)$ of the field $O_\chi$ in field theory. According to the dictionary, the retarded propagator is again the ratio of the leading boundary components, $\chi_-/\chi_+$, of the fluctuation $\Delta\chi(z;\omega,k)$ which satisfies exactly the same Klein-Gordon equation (\ref{scalareom}) as the equilibrium solution, only at finite energy and momentum:
\be
\label{gretchi}\delta\chi''+\left(\frac{f'}{f}-(D-1)A'-\frac{\partial_\Phi Z}{Z}\Phi'\right)\delta\chi'-\left(\frac{\omega^2}{f^2}-\frac{k^2}{f}+\frac{e^{-2A}}{f}m_\chi^2-\frac{q^2}{f^2}e^{\tau\Phi}A_0^2\right)\delta\chi=0.
\ee
Unlike the BF bound calculation, we are not exclusively interested in the case when the energy $\omega$ is pure imaginary but will consider general values of energy (with non-positive imaginary part, since the poles in the upper half-plane are forbidden).

%The conductivity for the non-condensed case a detailed analysis for a similar system was done in \cite{bigkir,goutkir}. There, the authors reformulate the fluctuation equation %(\ref{deltax}) as an effective Schr\"odinger problem and conclude, from the form of the effective potential, that the conductivity is gapless and follows a power-law scaling %$\sigma(\omega)\sim\omega^n$, roughly similar to a conductor. XXX In a neutral condensed system, XXX. Finally, the charged condensed case is a holographic superconductor and should %display a gap at small $\omega$.

\subsection{Effective Schr\"odinger equation for the response functions}

It is well-known (e.g.~\cite{zerotemp,mit}) that the IR behavior of the effective Schr\"odinger problem for various quantities like (\ref{deltax,deltat,gretchi}) is related to the energy scaling of the corresponding response functions in field theory, defined as the ratio of the leading and subleading component of the bulk field in the boundary. The aforementioned references study the case when the equation can be written in the form $A_x''-V(z)A_x=-\omega^2A_x$ (and similarly for any other field instead of $A_x$) with $V(z)\sim 1/z^2$ in the IR. The inverse-square potential is famous for allowing a conformal-invariant solution, and simple scaling arguments together with flux conservation lead to the conclusion that the $z$-scaling of the solutions to the Schr\"odinger equation in IR determines the $\omega$-scaling of the response function (essentially, the solution is a function of $\omega z$ only, and since the flux must be conserved ($z$-independent) it is also $\omega$-independent, which relates the scaling with $z$ to the scaling with $\omega$). In our problem, even in the deconfined case with no soft wall, the behavior of the potential is in general different from $1/z^2$, and no quantitative results on the frequency scaling can be drawn. We can, however, decide if the spectrum is gapped or continuous, and if the gaps are "hard" (zero spectral weight of the response function) or "soft" (exponentially suppressed nonzero weight).

As the charge susceptibility in dilaton systems was never studied so far, we give a more detailed analysis of the effective potential. The equation (\ref{deltat}) can be recast as a Schr\"odinger problem with an effective potential
\be
\label{veffchi}V_\mathrm{eff}(z;\omega,k)=-\frac{\omega^2}{f^2}+\frac{k^2}{f}+\frac{e^{-2A}}{f}m_\chi^2+\frac{X''}{X}+B\frac{X'}{X}
\ee
with $B=(D-3)A'-\tau\Phi'$ and $X=e^{-\int B/2}=e^{(D-1)/2A-\gamma\Phi}/\sqrt{f}$. Starting from the confined phase (in the charge-neutral case), we see that the potential for the confining geometry behaves in the IR as
\be
\label{veffia}V_\mathrm{eff}(z\to\infty;\omega,k)=-\omega^2+k^2-\frac{\nu(\nu-1(\tau-D+3))}{2}z^{\nu-2}+\frac{3}{4}\nu^2(\tau-D+3)^2z^{2\nu-2}+\ldots,
\ee
thus it grows to infinity in the IR (the subleading terms were left out). For finite $z$ (still far enough from the AdS boundary), it is positive if $\omega^2<\omega_0^2+k^2$ for some constant $\omega_0$, i.e. the spectrum is discrete and gapped for small energies. In the bulk, a gap in the spectrum simply means that there is no tunneling of the infalling solution toward the far IR at $z\to\infty$ (in the terminology of \cite{zerotemp}, the reflection coefficient is zero). This means that the integral $\int dz\sqrt{2V_\mathrm{eff}(z)}/z^2$ has to diverge at large $z$. For (\ref{veffia}) the integral behaves as $\int dzz^{\nu-3}$ and thus diverges for $\nu\geq 2$. Therefore, the gaps might be hard or soft depending on the parameters.

For $\omega>\sqrt{\omega_0^2+k^2}$ we expect a continuum, as the effective potential does not have a well anymore. In the deconfined neutral background, the potential looks like
\be
\label{veffibC}V_\mathrm{eff}(z\to\infty;\omega,k)=-\frac{(D-3)\alpha-\phi_0\tau}{2z}-\omega^2+k^2+\frac{3}{4z}((D-3)\alpha-\phi_0\tau)^2(\log z)^2+\ldots,
\ee
which grows to infinity in the IR but logarithmically slowly, whereas on the other side it again depends on $\omega-\sqrt{\omega_0^2+k^2}$. The spectrum is thus still gapped and discrete but (since the well is now shallow, because of the logarithmic growth) the bound states are expected to come closer to each other. Also, the tunneling probability behaves as $\int dz\log z/z^2$ which is finite for $z\to\infty$, and the gaps is always soft. In the charged case, the effective potential is augmented by a positive term proportional to $q^2A_0\chi^2$ which is independent of $\omega,k$. Therefore, the threshold $\omega_0$ is increased but the qualitative behavior remains the same. Similar conclusions hold for the other response functions: the gaps are always soft for the deconfined phase, and may be hard or soft for the confined phase.

\subsection{Numerics}

\subsubsection{AC conductivity}

\begin{figure}[ht!]
\begin{center}
(A)\includegraphics[width=0.45\linewidth]{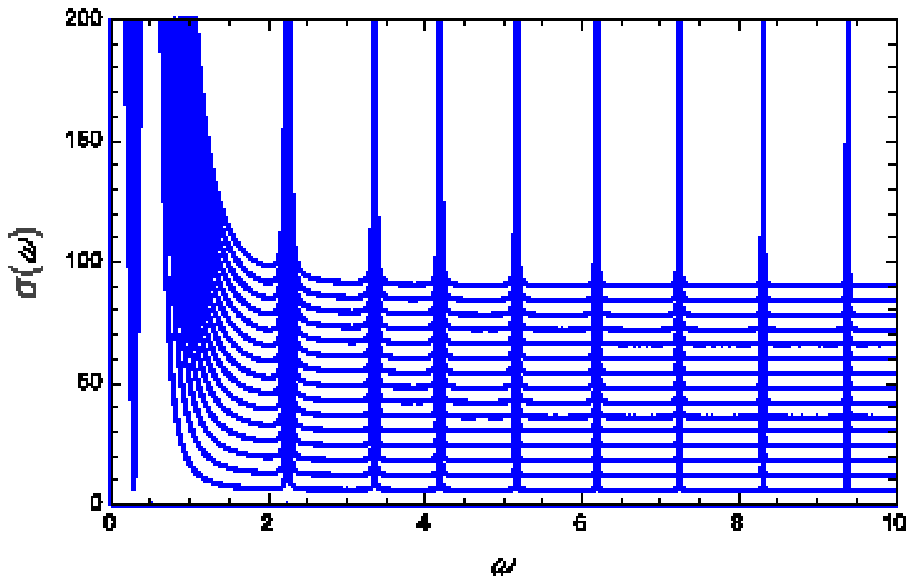} % SoloBar24a, BoseSpec24
(B)\includegraphics[width=0.45\linewidth]{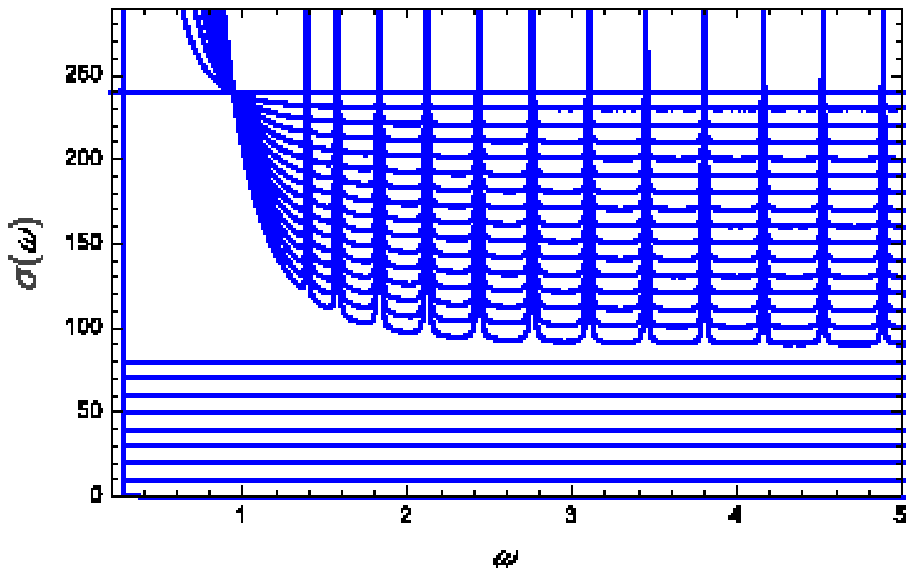} % SoloBar05, BoseSpec05a
(C)\includegraphics[width=0.45\linewidth]{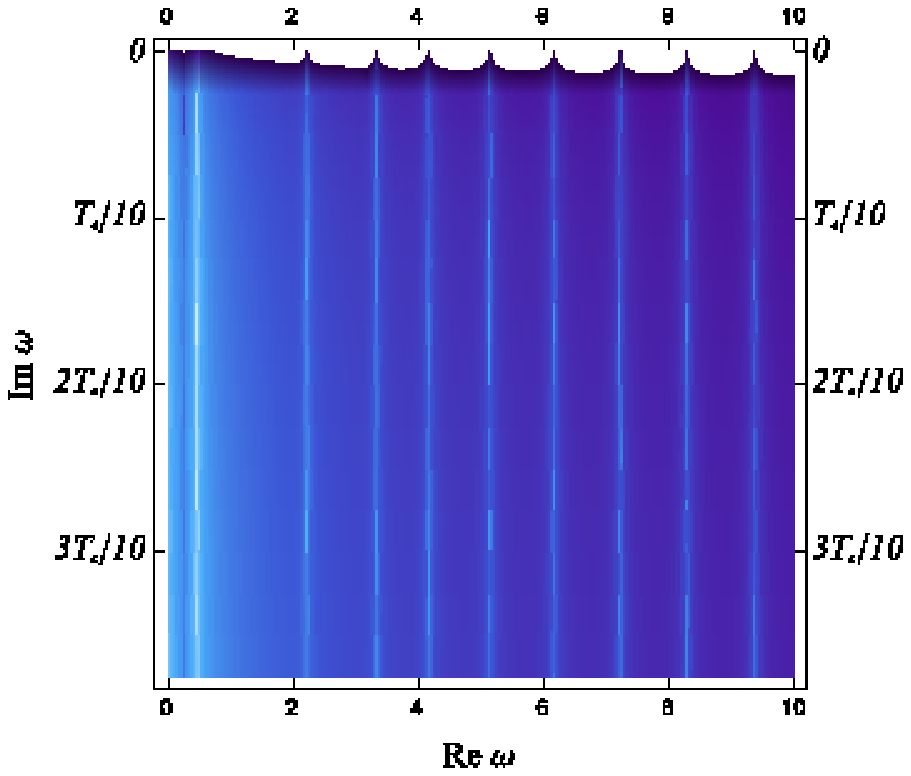}
(D)\includegraphics[width=0.45\linewidth]{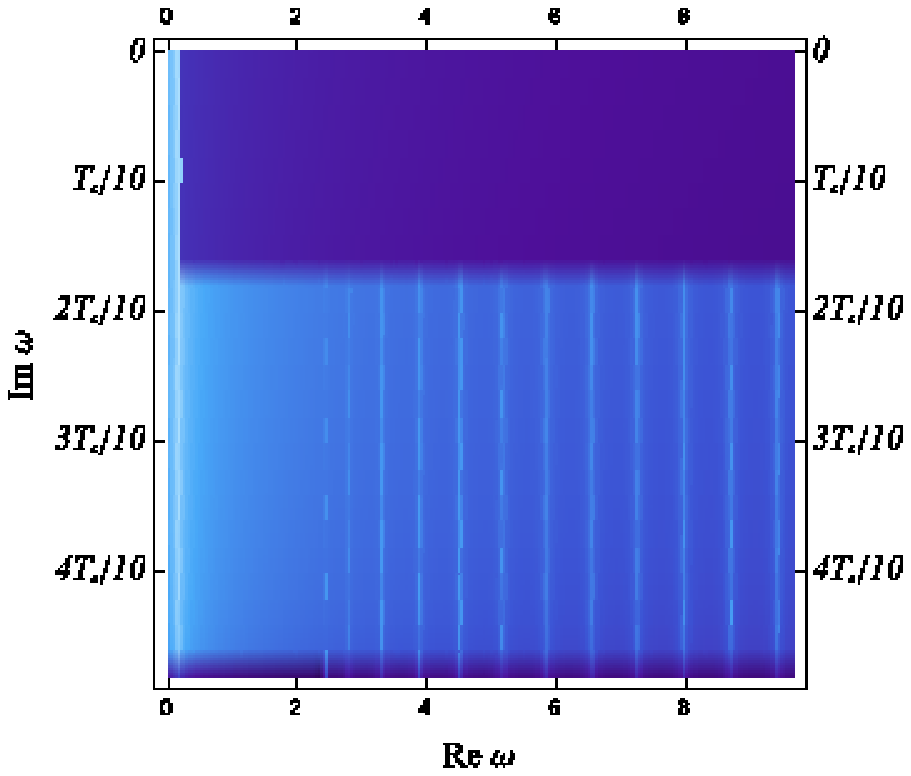}
\caption{Conductivity $\Re\sigma(\omega)$ in the confined/non-condensed phase ($m_\chi^2=2$, A,C) and in the deconfined/condensed phase ($m_\chi^2=-2$, B,D) in a $U(1)$-charged system at $\mu=1$, for a range of $\Im\omega$ values starting from zero (the real axis). In the deconfined/superconducting phase there is only the $\omega=0$ pole at the real axis (visible for the first curve in the panel B; in the color map panel D it is hard to recognize since it is very narrow), followed by a gap. The gap is expectedly absent in the confined/non-superconducting phase, as the continuous $U(1)$ symmetry is preserved. On the other hand, the confinement/deconfinement transition is visible through the stability of bound states: in the confined regime these states have an infinite lifetime at $T=0$ and thus manifest as sharp peaks (poles) on the real axis (the bright white spots on the real axis in the density plot). In the deconfined regime these states are pushed to a finite distance below the real axis and look more like branch cuts. For all calculations in a charged system in this section we use $D=4,\nu=3/2,\gamma=4,\tau=6,\mu=1$ and $m_\chi^2=1/4$ for the confined case and $m_\chi^2=-1/4$ for the deconfined case.}
\label{figsigmacharge}
\end{center}
\end{figure}

The AC conductivity best encapsulates the breaking of a continuous symmetry (\ref{symm}) through the existence of the zero mode. The AC conductivity on the real frequency axis, as well as in the bottom half-plane of complex $\omega$, is given in Fig.~\ref{figsigmacharge}. In this plot we show the conductivity $\Re\sigma(\omega,k=0)$ as a function of the real frequency $\Re\omega$ for a range of $\Im\omega$ values (at zero momentum). We first show the set of curves $\Re\sigma(\Re\omega)$ computed at different $\Im\omega$ values, where the curves at different $\Im\omega$ values are vertically shifted in the figure to be visible together (panels A, B); the $x$-axis is the real frequency axis and the $y$-axis is the magnitude of the conductivity minus the vertical shift. In parallel we show the same data as two-dimensional color maps $\Re\sigma(\Re\omega,\Im\omega)$ (panels C, D); now the $y$-axis is the imaginary part of the frequency, and the lighter areas denote higher values. We use the same recipe to show the curves $\Im G_R(\omega,k)$ and $\xi(\omega,k)$ in later figures.

In the charged confined non-condensed system (panels A, C), there is no gap at small frequencies as the continuous $U(1)$ symmetry is preserved. On the other hand, confinement means the existence of stable bound states ("glueballs"), i.e. poles on the real axis. These are seen as sharp peaks in $\Re\sigma(\omega)$ for real $\omega$. For nonzero $\Im\omega$ the poles apparently turn into branch cuts (the vertical lines); the resolution of our numerics is limited so we are not sure if these are branch cuts or strings of poles along the vertical ($\Im\omega$) axis. Such poles on the real axis have been seen also in \cite{horroberts} in the simple holographic superconductor (without dilaton) when the scalar mass is exactly at the BF bound for AdS${}_{D+1}$; the relation to our result is not clear but this fact is certainly interesting and we plan to look more carefully into it. Naively, it looks like a bad metal: the AC conductivity is continuous and gapless but small except on a discrete set of real frequencies where the bound states lie.

After deconfinement and the onset of superconductivity (Fig.~\ref{figsigmacharge}B, D), the Dirac delta peak at $\omega=0$ is followed by a gap, which shows the breaking of the $U(1)$ symmetry (this is particularly obvious in the panel B). The bound states do not sit at the real axis anymore. It is again not clear from the numerics if they turn into branch cuts or strings of poles in the complex plane but in any case they do not reach the real axis anymore. In this and further spectral plots, we use the critical temperature as a suitable unit of energy to express the frequencies and momenta; a more usual choice would be the chemical potential, but it is absent in the neutral case, so we have opted for $T_c$ as a natural and physical scale.

\begin{figure}[ht!]
\begin{center}
(A)\includegraphics[width=0.45\linewidth]{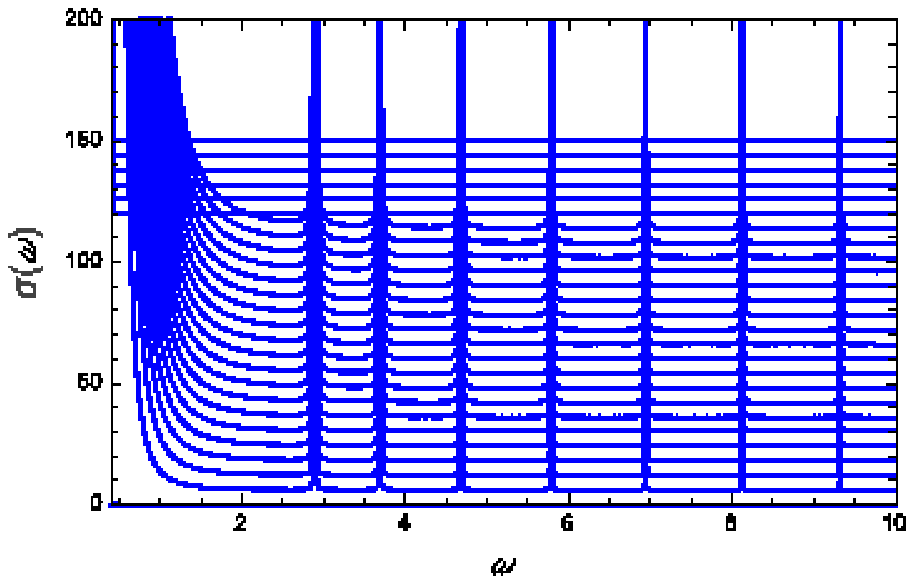} % SoloBar13, BoseSpec13
(B)\includegraphics[width=0.45\linewidth]{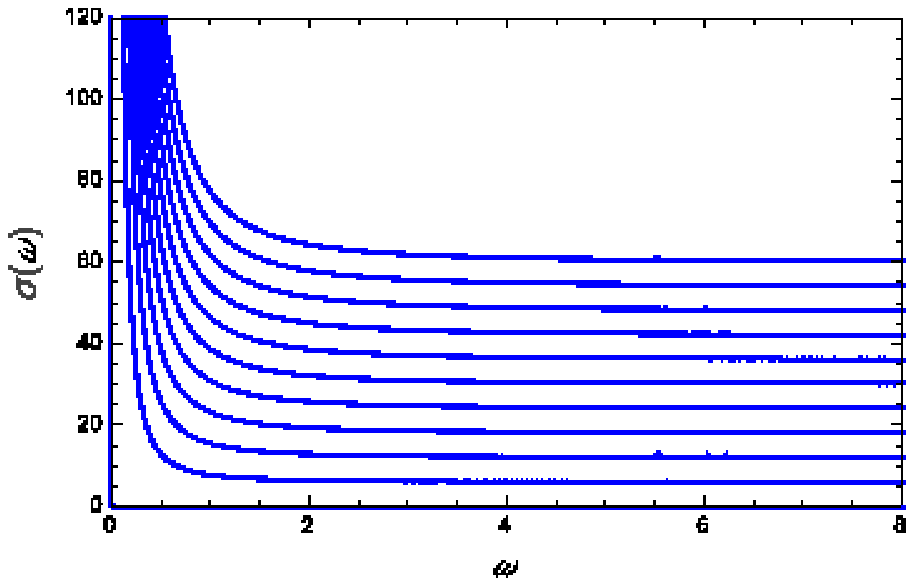} % SoloBar31, BoseSpec31
(C)\includegraphics[width=0.45\linewidth]{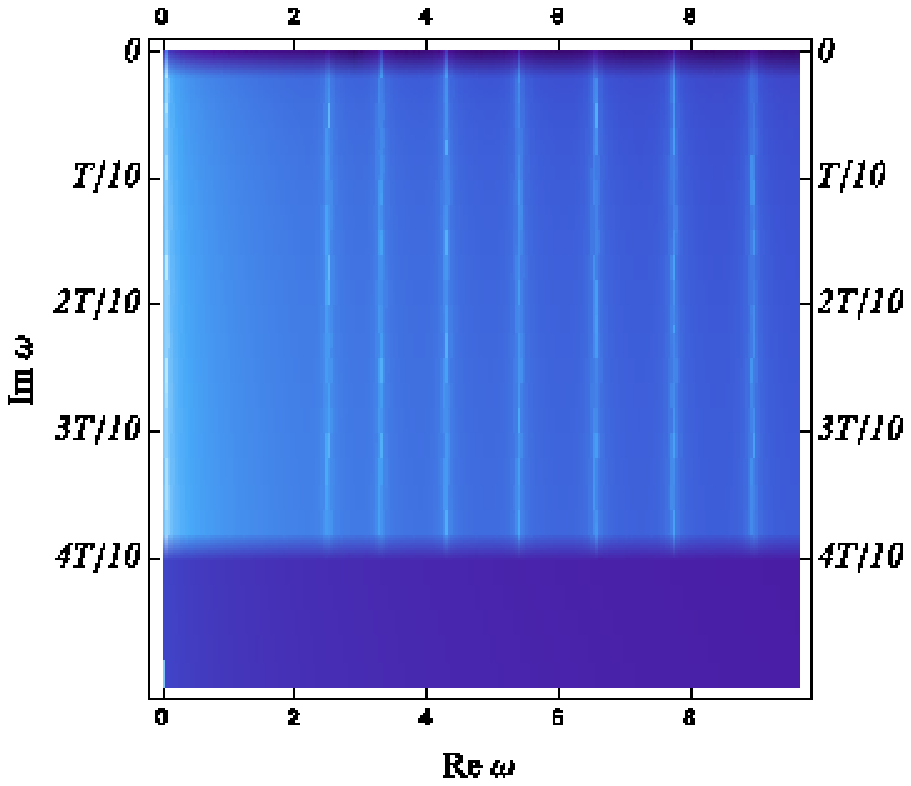}
(D)\includegraphics[width=0.45\linewidth]{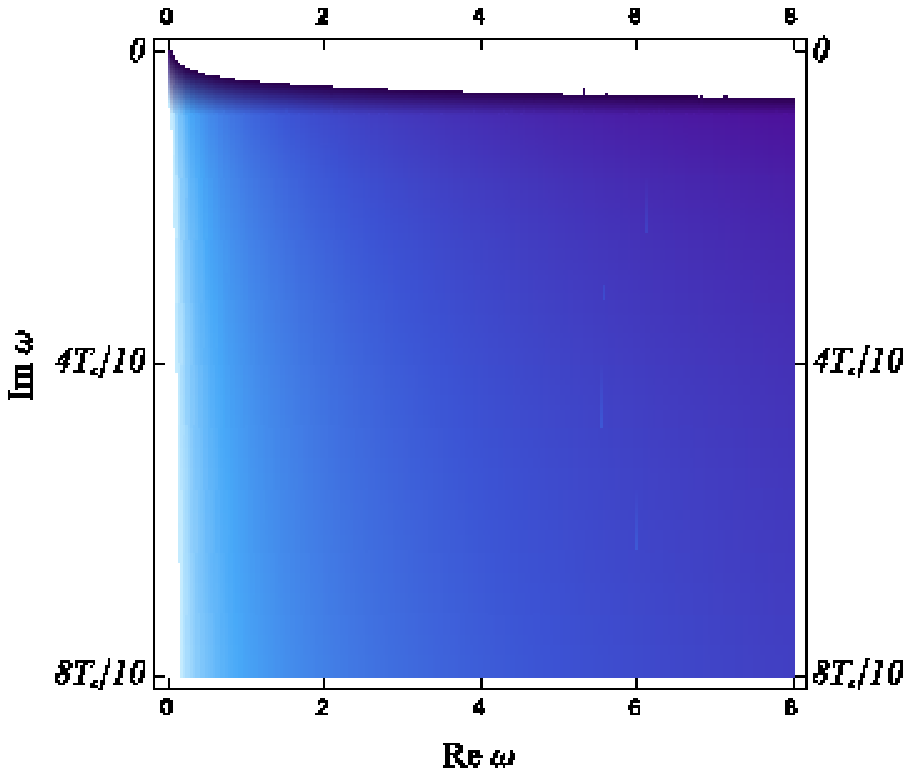}
\caption{Conductivity $\Re\sigma(\omega)$ in the confined/non-condensed phase ($m_\chi^2=8$, A,C) and in the deconfined/condensed phase ($m_\chi^2=4$, B,D) in a neutral system, for a range of $\Im\omega$ values starting from zero (the real axis). Neither phase is superconducting thus neither phase has a gap but rather a continuous background behaving as $1/\omega^n$. But the confined case again has long-living bound states corresponding to poles on the real axis, while upon deconfinement these poles vanish completely. The parameters are $D=4,\nu=3/2,\gamma=4$ and $m_\chi^2=4$ for the deconfined case and $m_\chi^2=8$ for the confined case (also in the remaining plots for the neutral system in this section).}
\label{figsigmaneut}
\end{center}
\end{figure}

Therefore, we witness both the breaking of the $U(1)$ symmetry (Dirac delta peak followed by the gap) and the deconfinement (absence of stable bound states), but not the restoration of scale invariance since our probe is charged and sees the nonzero chemical potential which sets a scale. It is instructive to compare this situation to the charge-neutral case in Fig.~\ref{figsigmaneut}. The superconducting gap now has to vanish from the spectrum. Only the presence or absence of confinement is now seen -- bound states as poles on the real axis (again apparently continuing as branch cuts below the real axis) in the confined regime and their absence in the deconfined regime. Notice that our confined phase is fractionalized and the deconfined phase is coherent -- therefore, our poles are \emph{not} "mesinos", they are closer to "glueballs", i.e. complex bound states which contain charged gauge bosons.

% figsigma3small, figsigma4small, figsigma3Sm, figsigma4Sm
% figsigma1small, figsigma2small, figsigma1Sm, figisgma2Sm
% figgr3small1, figgr4small, figgr3Sm1, figgr4Sm
% figgr1small, figgr2small, figgr1Sm, figgr2Sm
% figxi1small, figxi2small, figxi1Sm1, figxi2Sm

\subsubsection{Retarded propagator}

\begin{figure}[ht!]
\begin{center}
(A)\includegraphics[width=0.45\linewidth]{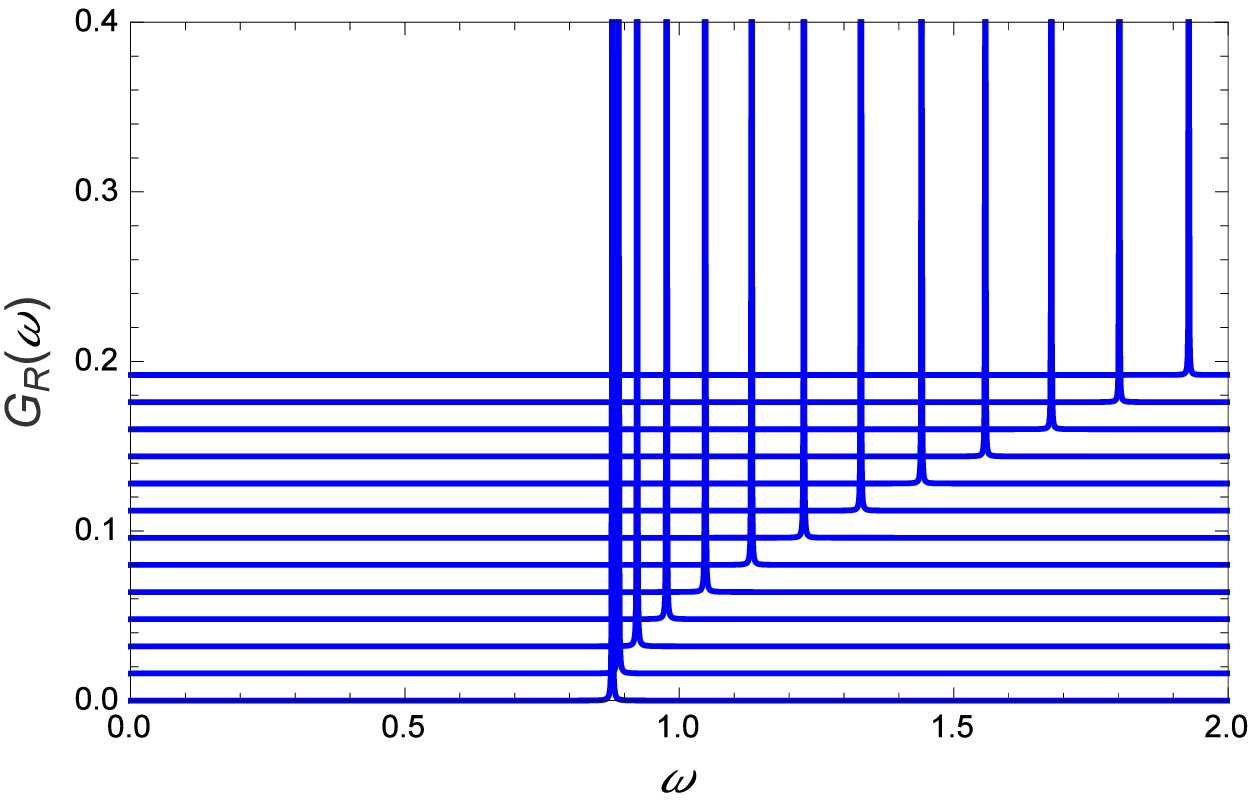} % SoloBar21, BoseSpec21
(B)\includegraphics[width=0.45\linewidth]{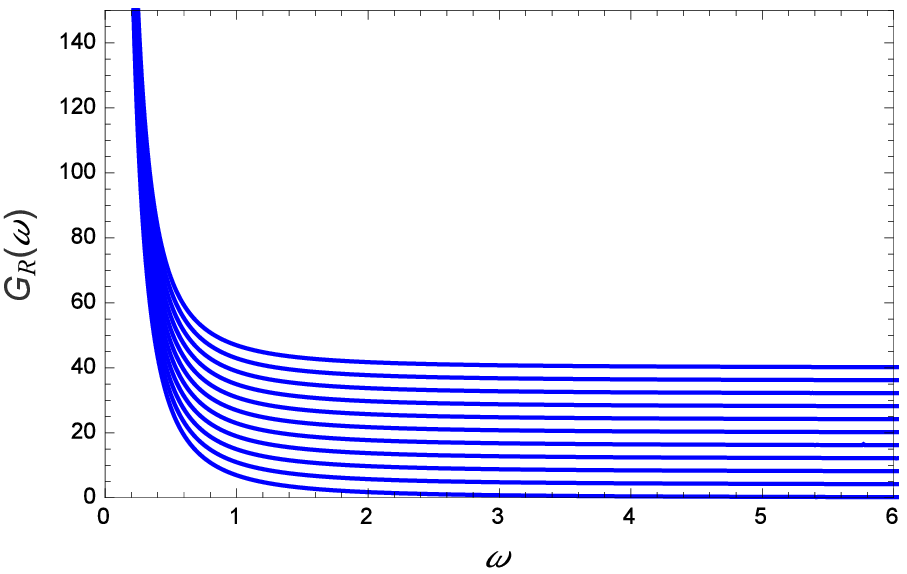} % SoloBar39; BoseSpec39
(C)\includegraphics[width=0.45\linewidth]{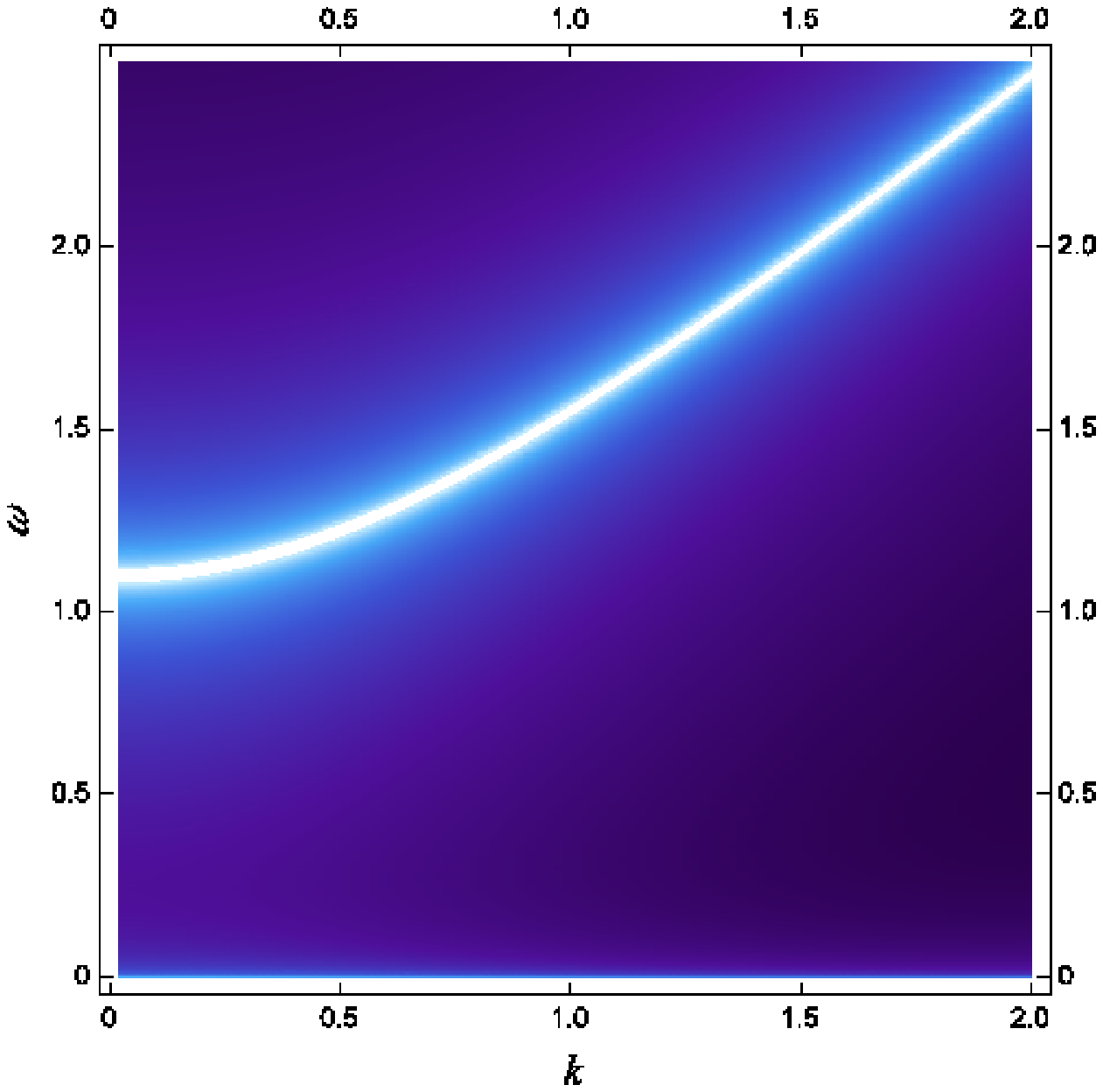}
(D)\includegraphics[width=0.45\linewidth]{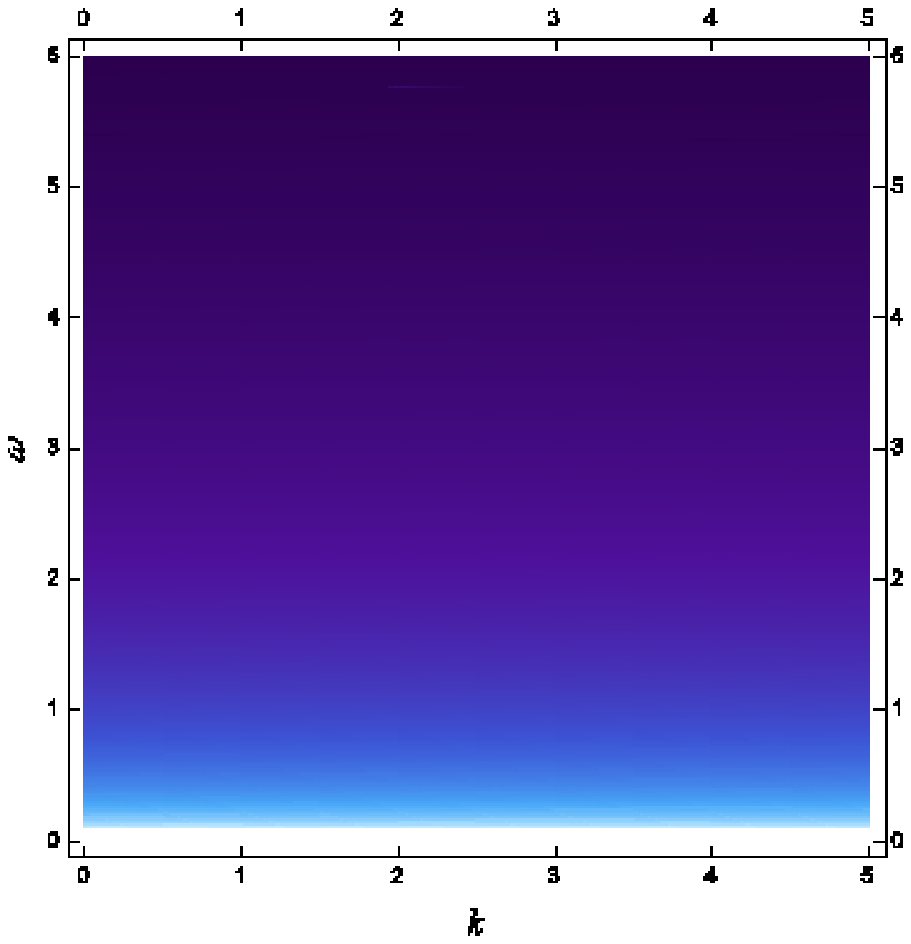}
\caption{The retarded propagator $\Im G_R(\omega)$ for a range of momentum values ($0<k<1.5$) in a charged system ($\mu=1$), in the confined regime ($m_\chi^2=-2$, A,C) and in the deconfined regime ($m_\chi^2=2$, B,D). In the confined case we see gapped quasiparticle excitations, starting at $\omega\approx 1>0$ since we see the bound states in the soft wall which have a discrete and gapped spectrum. In field theory, it means we see gauge-neutral particles. In the deconfined regime, no quasiparticle is present and we have a featureless power-law spectrum $\Im G_R(\omega)\propto1/\omega^n$. From the gravity viewpoint, it is because the potential has no bound states. From the field theory viewpoint, it means we have gauge-colored excitations which are not visible through a gauge-neutral probe. We thus see the deconfinement transition.}
\label{figgrcharge}
\end{center}
\end{figure}

\begin{figure}[ht!]
\begin{center}
(A)\includegraphics[width=0.45\linewidth]{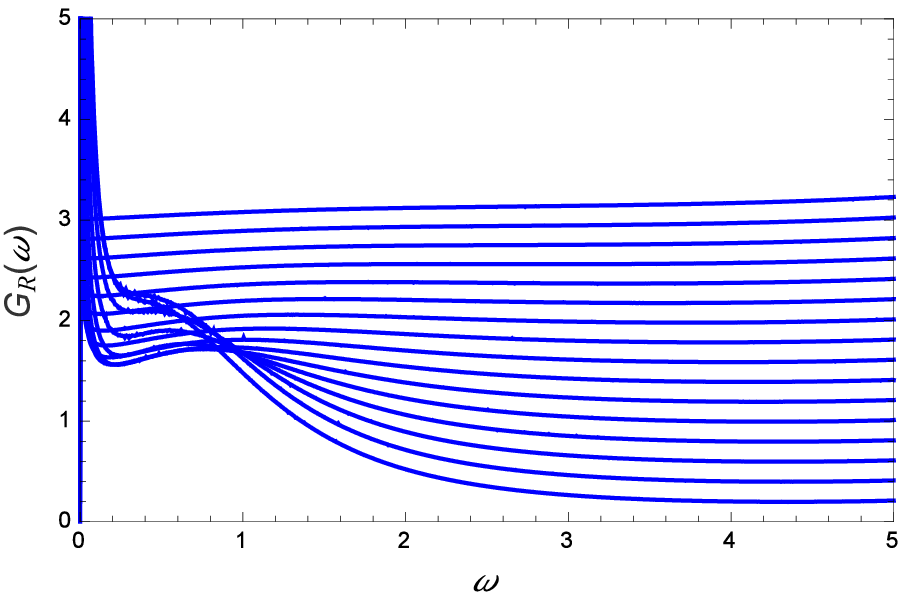} % SoloBarMap12, BoseSpec12
(B)\includegraphics[width=0.45\linewidth]{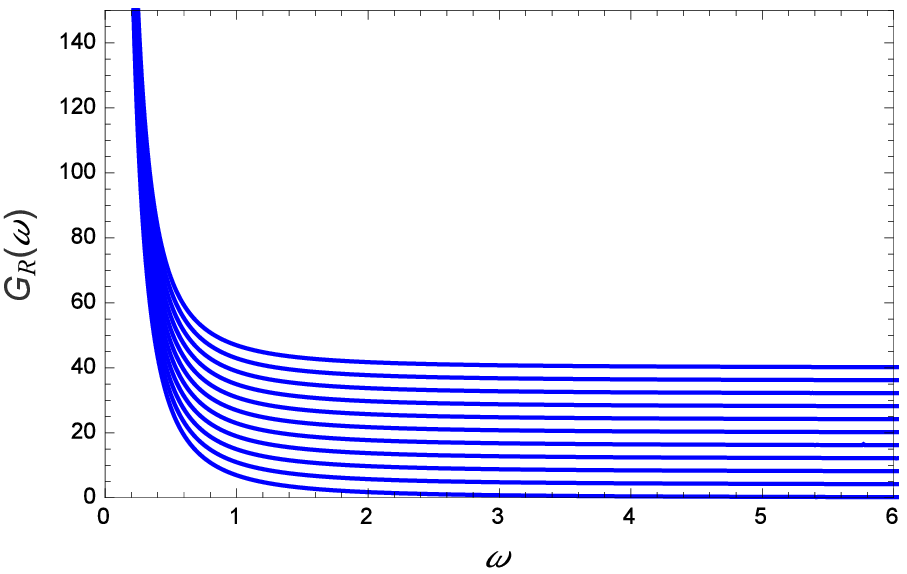} % 39a
(C)\includegraphics[width=0.45\linewidth]{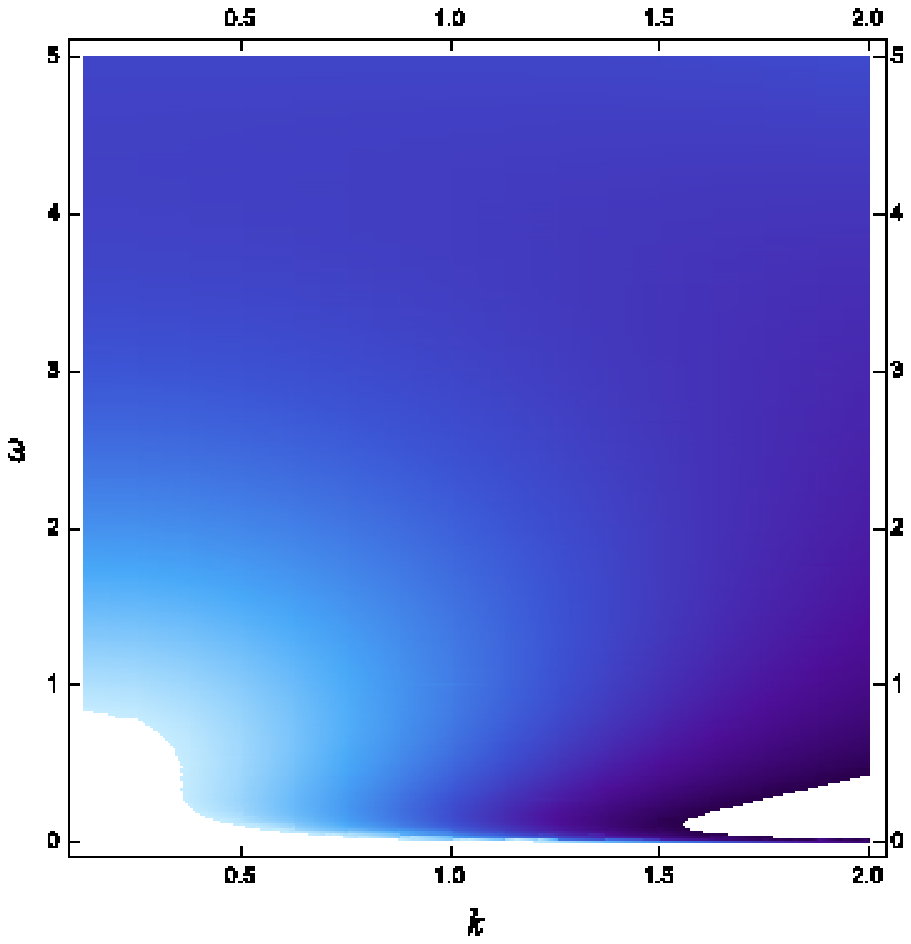}
(D)\includegraphics[width=0.45\linewidth]{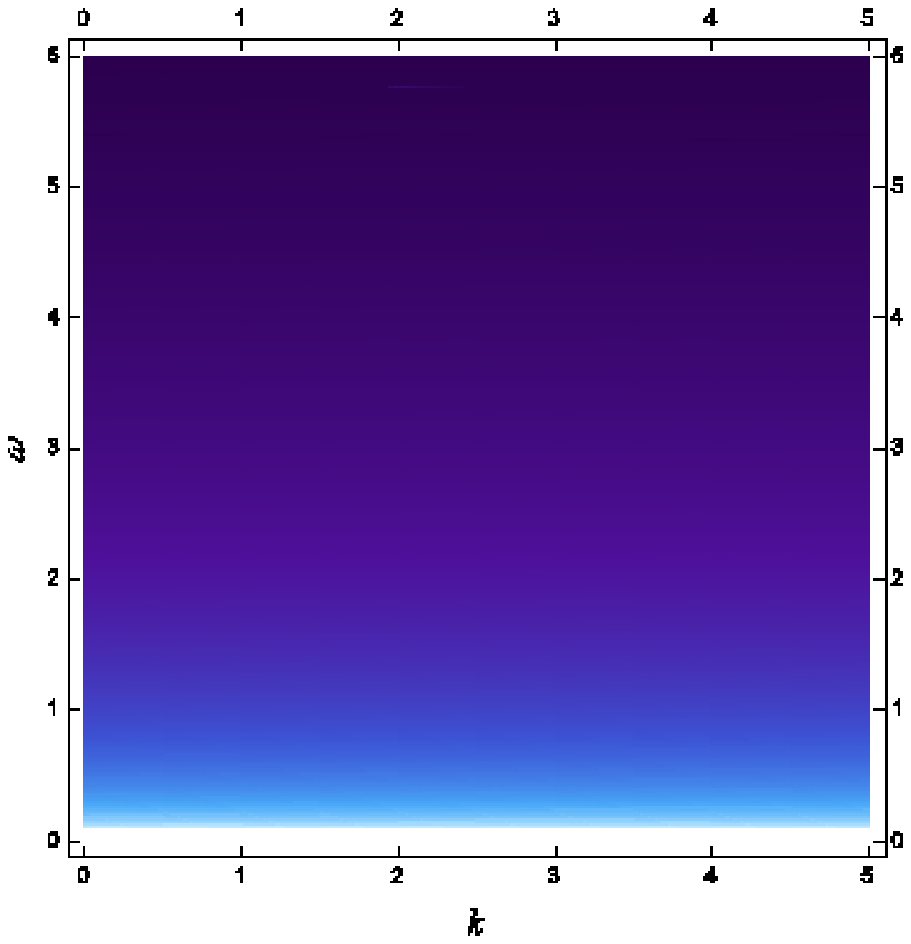}
\caption{The retarded propagator $\Im G_R(\omega)$ in the confined/non-condensed phase (A,C) and in the deconfined/condensed phase (B,D) in a neutral system. While the deconfined case is again an almost exact power law, the confined case has a scale but no quasiparticle. The discrete $Z_2$ symmetry has no zero modes upon breaking. The retarded propagator is thus not so useful when the system is neutral.}
\label{figgrneut}
\end{center}
\end{figure}

A probe which specifically shows the restoration of scale invariance is the retarded propagator $G_R(\omega,k)$, given in Fig.~\ref{figgrcharge}. In the confined regime we see well-defined quasiparticles, due to nonzero chemical potential. But since quasiparticles exist at finite binding energies, the spectrum is gapped and starts from nonzero energy (A, C). Once the system is deconfined, scale invariance is restored and $G_R(\omega)\sim 1/\omega^n$ (B, D). Unlike conductivity, which is not sensitive to confinement/deconfinement, the scalar probe differentiates between them: in their absence, it shows no quasiparticles. Another way to understand it is that at low energies (in deep interior) the local chemical potential behaves as $e^{-A}A_0/\sqrt{f}\sim z^{-2\alpha+\phi_0\tau/2}$ while the scale of the metric (the confinement scale) drops faster as $z^{-2\alpha}$, so the confinement scale is above the chemical potential and the probe sees no chemical potential at all. When the system is neutral and the symmetry to be broken is discrete, we expect to see the presence/absence of scale invariance in much the same way as before but we expect no quasiparticle in either phase, since the chemical potential is zero. The plot for the neutral case is shown in Fig.~\ref{figgrneut}: now there is indeed no quasiparticle in either phase as the chemical potential and density are zero. But we still detect a scaleful, though continuous spectrum in the confined case, whereas the deconfined case looks pretty much the same as with a charged boson -- just a power-law decay. Again, this is \emph{not} about fractionalization -- the confined phase, with quasiparticles in Fig.~\ref{figgrcharge}(A,C), has "gauginos" which the gauge-neutral probe cannot see. and the deconfined phase, with no quasiparticles in Fig.~\ref{figgrcharge}(B,D), has "mesinos" which the gauge probe can see. The bottom line is that the probe apparently couples mainly to the gauge field bound states, and in general that the presence/absence of quasiparticles may not be directly related to fractionalization.

\subsubsection{Charge susceptibility}

\begin{figure}[ht!]
\begin{center}
(A)\includegraphics[width=0.45\linewidth]{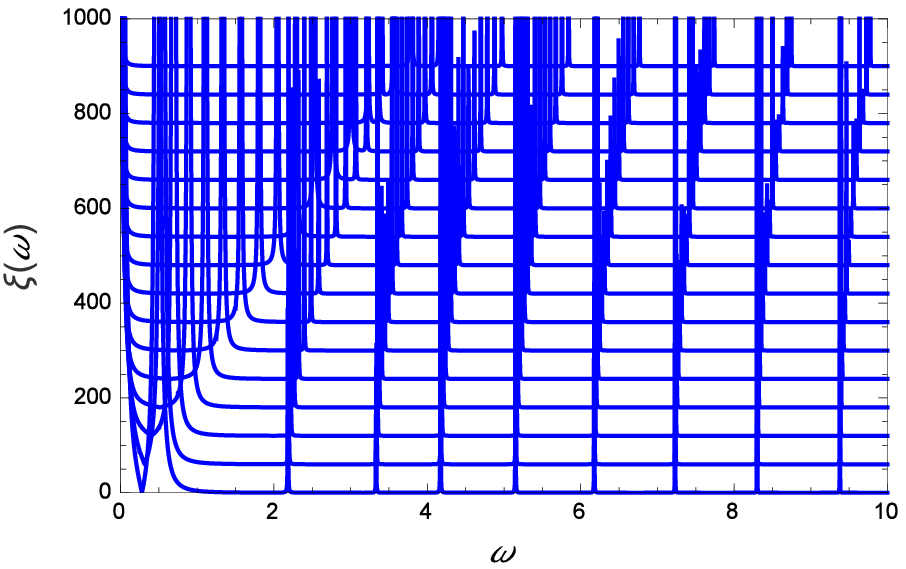} % SoloBar23
(B)\includegraphics[width=0.45\linewidth]{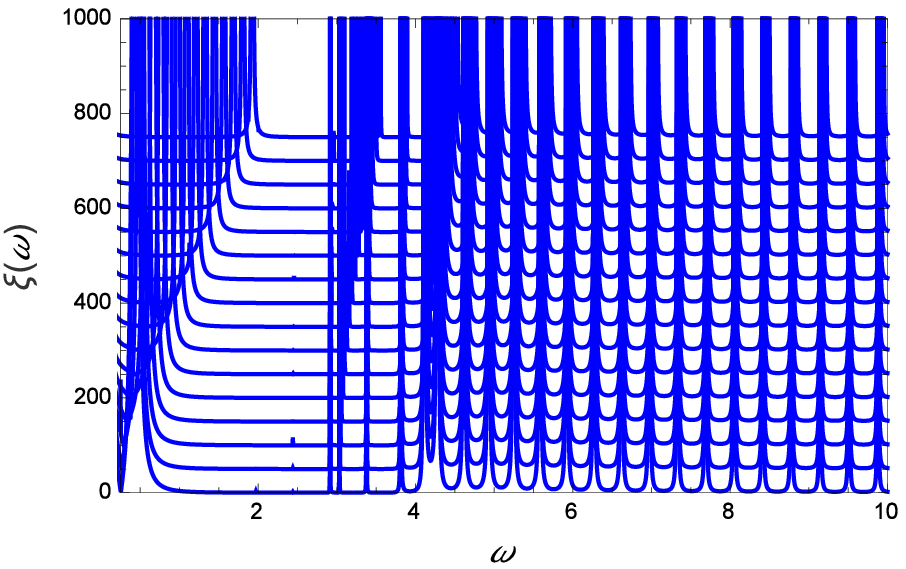} % SoloBar15
(C)\includegraphics[width=0.45\linewidth]{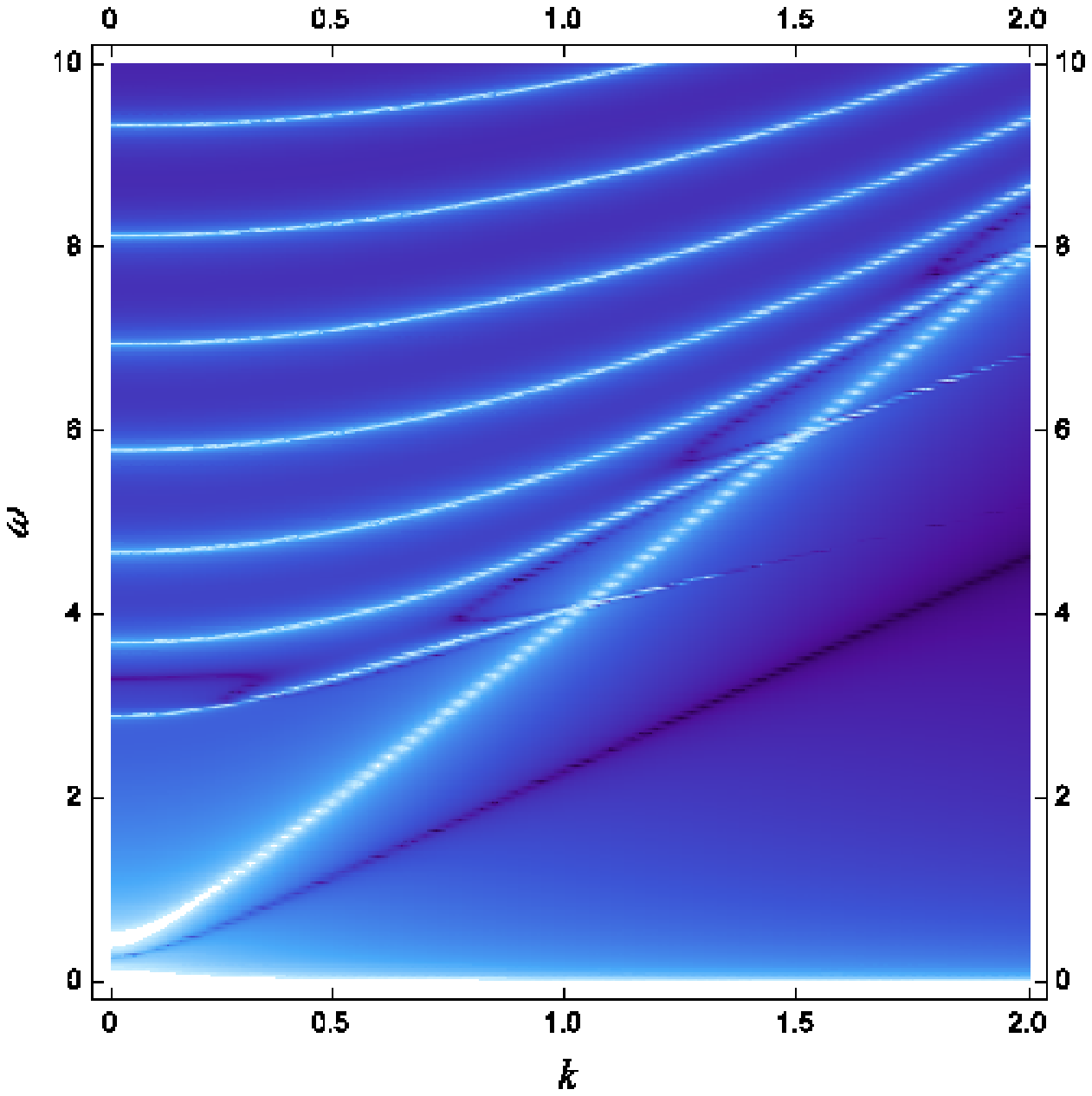}
(D)\includegraphics[width=0.45\linewidth]{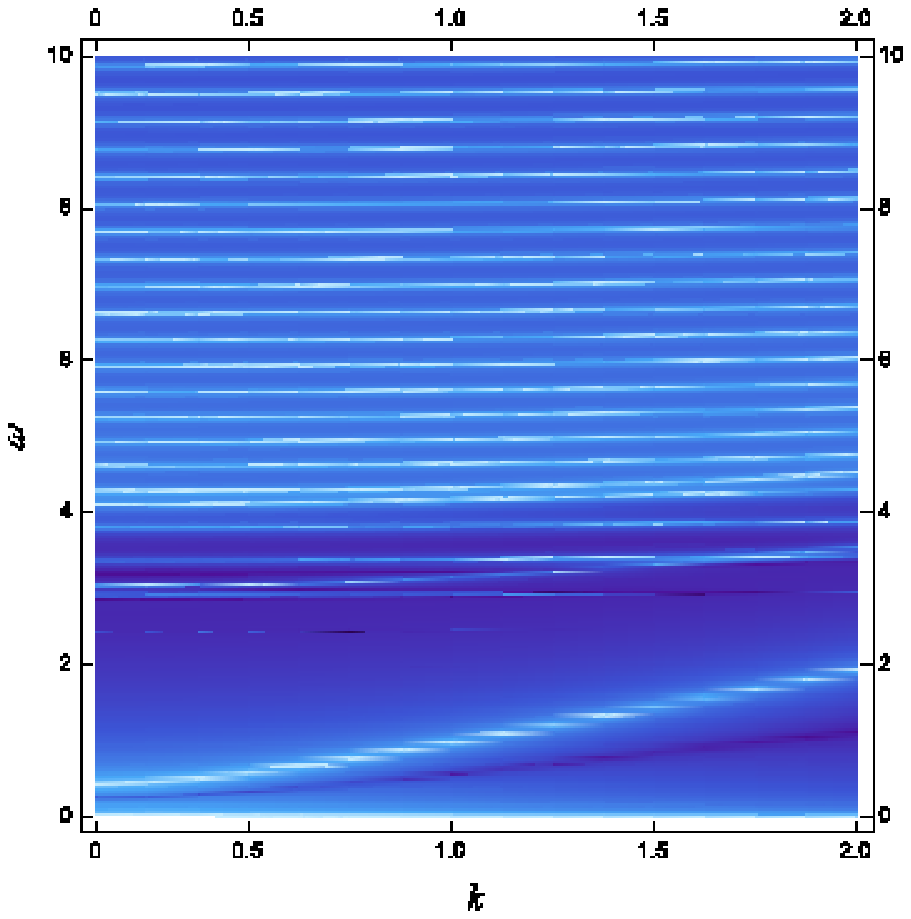}
\caption{Charge susceptibility $\xi(\omega)$ in the confined/non-condensed phase (A,C) and in the deconfined/condensed phase (B,D) in a charged system. Both cases show bound states; this might look surprising for the deconfined case but is in accordance with the effective potentials in Eqs.~(\ref{veffia}-\ref{veffibC}). This probe is thus not very useful for detecting the transition but shows the absence of peaks at $\omega=0$ and $k=2k_F>0$, indicating that even the confined phase is different from a normal Fermi liquid.}
\label{figxicharge}
\end{center}
\end{figure}

Charge susceptibility is interesting as it shows the absence of metallic behavior in both the confined and deconfined phase. Both phases show a gap followed by a series of dispersing poles. This is in line with our analytical finding that both backgrounds give a potential well for $\delta A_0$, inhabited with bound states. But since the well is rising towards infinity very slowly in the deconfined phase, the spacing between the bound states is small in this case. In \cite{donos} the authors have explored mainly the momentum dependence of the susceptibility at zero frequency, finding the Friedel oscillations and the singularity at $k=2k_F$, as expected for a system with zero modes at \emph{finite} momentum, resembling a Fermi surface. In Fig.~\ref{figxicharge}, in particular in the $\omega-k$ maps (panels B,D) we see that no oscillatory behavior exists for $\chi(\omega=0,k)$ (the bottom edge of Fig.~\ref{figxicharge} C,D) and in particular no pole at $\omega=0$ exists for any finite $k$. This tells that our system is different from a normal metal even in the confined phase, and this is not because it is fractionalized (since the RN black hole studied in \cite{donos} is also fractionalized).

\section{Conclusions and discussion}

We have considered an Einstein-(Maxwell)-dilaton-scalar system where the scalar can condense (acquire a VEV) and thus break a symmetry, discrete if neutral or continuous if charged. This in turn remorphs the geometry from a soft-wall, confining form to a deconfined, power-law-scaling form. This goes against the common intuition that a condensate always "narrows" the geometry, which indeed happens in absence of a dilaton with a suitably chosen coupling, e.g. in the textbook holographic superconductor where an AdS-RN background with a near-horizon AdS${}_2$ throat with finite AdS radius typically turns into a Lifshitz-type geometry whose scale shrinks to zero in the interior. From a general viewpoint, it is not so surprising that the huge "zoo" of dilatonic theories contains counterexamples to this behavior, as we have great freedom in choosing the dilaton potentials. But from the viewpoint of field theory and applied gauge/gravity duality, this is interesting as it tells us that we can consider situations in which breaking a symmetry with an order parameter can actually restore another symmetry, since confined systems have a scale (the confinement gap) which vanishes upon condensation. In the simplest case, we can thus expect that conformal symmetry is restored. In practice, it is not the full conformal symmetry but some subset of it, i.e. some scale invariance. We therefore see a non-Ginzburg-Landau phase transition, where neither phase has a higher overall symmetry than the other and the transition can be continuous (in the charged case). This may be related to the picture of deconfined criticality proposed in \cite{deconfcrit}. But one should be careful, since the transition mechanism in \cite{deconfcrit} is related to the existence of a new, topological conserved quantity which only exists at the critical point. In our setup we cannot study geometry or lattice effects and definitely cannot argue anything about topology. The connection is thus very loose and we only see it as inspiration for further work. It would be interesting to consider a setup where the topologically protected gauge flux analogous to that at a deconfined critical point can be detected.

In would also be nice to understand our system better from the gravity side, by deriving our solutions from a superpotential and inspecting how generic this behavior is, which we address in a subsequent publication. It is also interesting to apply our findings to real-world systems. While in QCD there is no obvious additional order parameter that may condense, such situations are abundant in condensed matter systems, mainly in the context of the fractionalization paradigm, where certain non-Fermi-liquid phases are argued to consist of gauge-charged excitations which are therefore not observable as quasiparticles. This is also relevant for the heavy fermion systems, where a long-range order is present (the antiferromagnetic ordering, the $SO(3)$ equivalent of our scalar neutral order parameter) and is connected to the disappearance of a normal Fermi liquid, which can be related to the deconfinement of the gauge-charged spinons and holons (in this case, of course, the gauge field is emergent and not microscopic) \cite{heavy}. However, great care must be taken to interpret the fractionalization concept properly, as it is distinct from confinement -- in our case, the confined phase is fractionalized and the deconfined phase is coherent. This will also be addressed in our future work.

\section*{Acknowledgments}

We are grateful to A.~Rosch for helpful discussions and careful reading of the manuscript as well as to M.~V.~Medvedyeva, K.~E.~Schalm and J.~Zaanen for helpful discussions and support.

\appendix

\section{A short summary of numerical calculations}
% App A

For numerical work we find it more convenient to introduce a different coordinate choice where the metric reads
\be
ds^2=-\frac{f(z)h(z)dt^2}{z^2}+\frac{d\mathbf{x}^2}{z^2}+\frac{dz^2}{f(z)z^2}.
\ee
The boundary is again at $z=0$ and the space extends to $z\to\infty$. It is easy to derive the relations between this parametrization and the one used in the main text. Now the boundary conditions for small $z$ are $f(z\to 0),h(z\to 0)\to 1$. The Einstein equations read
\bea
zf'-Df+D+\frac{2}{D-1}T^{00}=0\\
\frac{h'}{h}zf=\frac{2}{D-2}(T^{00}-T^{zz}).
\eea
Therefore, both metric functions have first-order equations and we can only impose two boundary conditions for the metric. However, we have more than two physical requirements. The physical requirement for $h$ (which is proportional to the scale factor $e^{-2A}$ in the metric (\ref{metricir})) is $h(z\to\infty)\to 0$ and for $f$ the first derivative should vanish: $f'(z\to\infty)\to 0$. In addition, in order to have an asymptotically AdS geometry we need $f(z\to 0)=1$ and $h(z\to 0)=1$. We implement this by introducing some cutoff $z_\Lambda$ and imposing the analytical solutions we have found for the metric in section III for all $z<z_\Lambda$ (the analytical solutions of course automatically satisfy the necessary requirements in the interior). Then we start the integrator at $z_\Lambda$, using the condition $f(z=0)=h(z=0)=1$ as the sole boundary condition for the numerics. At finite temperature, the space terminates at the horizon $z_h$ whose value is determined by the temperature, and in this case $f$ itself vanishes at the horizon: $f(z_h)=0$, whereas the derivative is known analytically: $f'(z_h)=\mathrm{const.}\times T$ (the constant can be calculated). In practice, it means we use the analytical ansatz for $f,h$ in the interval $z_h>z>z_h-\epsilon$ and start the integration at $z=z_h-\epsilon$, again with the boundary condition $h(z=0)=1$.

The equations of motion for the gauge and matter fields are
\bea
\Phi''-\left(\frac{h'}{2h}-(D-1)z\right)\Phi'+\frac{g_{zz}}{\xi}\partial_\Phi V-\frac{4}{\xi(D-1)}g^{00}\mathcal{T}A_0^2=0\\
A_0''-\left(\frac{h'}{2h}-(D-3)z-\frac{\partial_\Phi\mathcal{T}}{\mathcal{T}}\Phi'\right)A_0'-2q^2\frac{Z}{\mathcal{T}}\sqrt{g_{00}}g_{zz}\vert\chi\vert^2=0\\
\chi''-\left(\frac{h'}{2h}-(D-1)z\right)\chi'-\frac{m_\chi^2}{\xi z^2f}+\frac{q^2Z}{f^2}A_0^2\chi=0.
\eea
Here we have three second-order equations and two boundary conditions per field. For $A_0$, one condition is that the electric field should vanish in the interior: $-A_0'(z\to\infty)\to 0$ and the other is to impose the chemical potential or the charge density at the boundary ($A_0(z\to 0)=\mu$ or $A_0(z)/z^{D-2}\vert_{z\to 0}=-\rho$). For $\Phi$ and $\chi$ the only physically obvious boundary condition is to set the leading branch in the small-$z$ expansion (\ref{chibnd}) to zero (remember we pick the dilaton potential $V$ in the UV in such a way that the subleading branch of the dilaton also falls off quickly enough that no condensation occurs). The other boundary condition for $\Phi,\chi$ is again set by the analytical expansion for $z$ large, similar as for the metric.

It is well known that the integration is unstable if started from the boundary. We therefore start from the interior and impose all boundary conditions in the interior. Physical requirements for $z\to 0$ are then obtained by shooting. We start from $z_1\equiv z_\Lambda$ at $T=0$ or from $z_1\equiv z_h-\epsilon$ at finite $T$ and iterate the procedure in two stages. The first iteration assumes some essentially arbitrary metric in the whole space (AdS$_{D+1}$ works well) and solves first the coupled system for $f,h,A_0,\Phi$. For $f$, the boundary condition is the analytical estimate $f_\mathrm{anal}(z_1)$. We similarly impose the analytical estimate for $\Phi$ while for $\Phi'$ we try an arbitrary value $C_1$. For $h$ we also start from an arbitrary value $C_2$. For the gauge field we impose the physical boundary condition for the derivative ($A_0'(z_1)=0$) whereas the other condition is arbitrary ($A_0(z_1)=C_3$). We thus have three free parameters $C_1,C_2,C_3$ so we can shoot for the correct UV behavior of $A_0,\Phi,h$. This procedure does \emph{not} guarantee the correct behavior for $f(0)$ and $h(z_1)$ as we do not shoot for them but when one lands at the correct solution, these turn out to be automatically satisfied (if not, one should play around a bit with the starting values of the shooting parameter $h(z_1)$). In the next stage, we solve the equation for $\chi$ with the conditions $\chi(z_1)=C_4\chi_\mathrm{anal}(z_1)$ and $\chi'(z_1)=C_4\chi'_\mathrm{anal}(z_1)$, leaving the overall normalization $C_4$ as a free parameter. Then we shoot for the required behavior in the UV (this will yield the solution with nonzero VEV, if it exists; if not, it will give the solution $\chi(z)=0$). After that, we update the metric and the stress tensor and repeat the whole procedure, again in two steps, first for $f,h,\Phi,A_0$ and then for $\chi$. After $5-10$ steps (a few minutes of computation time) the procedure converges. One should check that the solution is independent of the cutoff $z_1$. At zero temperature, for confining backgrounds the overall scale falls off very sharply and typically $z_1\approx 3-4$ is enough while for nonconfining geometries one needs $z_\Lambda\approx 6-10$. At finite temperature, the size of the "analytical" region in the interior $\epsilon$ can be made quite small, of the order $10^{-3}$. A cutoff in the UV is also necessary and is roughly of the size $10^{-6}$.

\end{document}